\documentclass[acmlarge]{acmart}

\AtBeginDocument{%
  \providecommand\BibTeX{{%
    \normalfont B\kern-0.5em{\scshape i\kern-0.25em b}\kern-0.8em\TeX}}}

\usepackage{comment}

\begin{document}

\title{Playing games with Tito: Designing hybrid museum experiences for critical play}

\author{Anders Sundnes Løvlie}
\email{asun@itu.dk}
\orcid{0000-0003-0484-4668}
\affiliation{%
  \institution{Center for Computer Games Research, IT University of Copenhagen}
  \streetaddress{Rued Langgaards Vej 7 }
  \city{København}
  \postcode{2300, Denmark}
}
\author{Karin Ryding}
\affiliation{%
  \institution{Center for Computer Games Research, IT University of Copenhagen}
  \streetaddress{Rued Langgaards Vej 7 }
  \city{København}
  \postcode{2300, Denmark}
}
\author{Jocelyn Spence}
\affiliation{%
  \institution{Mixed Reality Lab, University of Nottingham}
}
\author{Paulina Rajkowska}
\affiliation{%
  \institution{Department of Informatics and Media, Uppsala University}
  \streetaddress{Box 513}
  \city{Uppsala}
  \postcode{751 20, Sweden}
}
\author{Annika Waern}
\affiliation{%
  \institution{Department of Informatics and Media, Uppsala University}
  \streetaddress{Box 513}
  \city{Uppsala}
  \postcode{751 20, Sweden}
}
\author{Tim Wray}
\affiliation{%
  \institution{Center for Computer Games Research, IT University of Copenhagen}
  \streetaddress{Rued Langgaards Vej 7 }
  \city{København}
  \postcode{2300, Denmark}
}
\author{Steve Benford}
\affiliation{%
  \institution{Mixed Reality Lab, University of Nottingham}
}
\author{William Preston}
\affiliation{%
  \institution{Mixed Reality Lab, University of Nottingham}
}
\author{Emily Clare-Thorn}
\affiliation{%
  \institution{Mixed Reality Lab, University of Nottingham}
}

\renewcommand{\shortauthors}{Løvlie et al.}

\begin{abstract}
This article brings together two distinct, but related perspectives on playful museum experiences: Critical play and hybrid design. The article explores the challenges involved in combining these two perspectives, through the design of two hybrid museum experiences that aimed to facilitate critical play with/in the collections of the Museum of Yugoslavia and the highly contested heritage they represent. Based on reflections from the design process as well as feedback from test users we describe a series of challenges: Challenging the norms of visitor behaviour, challenging the role of the artefact, and challenging the curatorial authority. In conclusion we outline some possible design strategies to address these challenges.

\end{abstract}

\setcopyright{acmlicensed}
\acmJournal{JOCCH}
\acmYear{2021} \acmVolume{1} \acmNumber{1} \acmArticle{1} \acmMonth{1} \acmPrice{15.00}\acmDOI{10.1145/3446620}





\maketitle

\section{Introduction}
There is a growing interest in play and games in museums, causing the museum scholar Jenny Kidd to declare a "ludic turn within museums" \cite{kidd_immersive_2018}. Games have been created for museums for a range of different purposes, such as increasing visitor engagement \cite{ali_deepening_2018, ciolfi_designing_2012}, learning \cite{anderson_developing_2010,malegiannaki_analyzing_2017,paliokas_sylaiou}, or community building \cite{holloway-attaway_performing_2020}. However, bearing in mind the educational role of museums, the turn towards playfulness may also raise concern: Can museums at the same time facilitate play and critical thinking? Kidd suggests that playful and immersive heritage encounters "can be a way of asking difficult questions and offering provocations on the very nature of museum-making" \cite{kidd_immersive_2018}.

This is part of the vision of critical play: play which purposely challenges dominant worldviews and systems of power \cite{flanagan_critical_2009}. Considering the concerns of new museology \cite{vergo_new_1989} and critical museology \cite{Shelton2013}, which combine an interest in participation and interactivity with a critical stance towards established systems of thought, critical play is a highly relevant perspective to explore for the design of playful museum experiences.

The variety of technologies used for museum experiences is ever increasing. Much research in Human-Computer Interaction (HCI) has explored the design of interactive technologies for museums, both considering standalone installations, smartphone-based experiences and various types of assemblages of devices \cite{hornecker_ciolfi_2019}. The perspective of ‘hybrid design’ \cite{bannon_hybrid_2005} focuses on designing experiences that merge the physical visit with digital content in new ways. Hybrid design has been used to enable participation \cite {claisse_augmented_2017, dalsgaard_designing_2008}, and in various ways provoke and challenge the visitor in order to invite reflection \cite{avram_co-designing_2016, ciolfi_including_2008, fosh_supporting_2016, hornecker_-and-fro_2016}. Smartphone technology makes it possible to create pervasive games \cite{montola_pervasive_2009}: Games that challenge the spatial, temporal and social limits for play. Thus, smartphone-based games are a particularly promising technology for critical play in the museum context, as they make it possible to use the museum as a hybrid physical/virtual playground, challenging not just the limits of play but also the norms and expectations of the museum as an institution.

Such endeavours are not free of obstacles, not least due to the complexity of combining new designs and technological formats with the critical stance. In this paper we report on our experiences developing and testing two hybrid museum experiences 'in the wild', \cite{benford_performance-led_2013, crabtree_introduction_2013}. A museum, a game design company and a research team collaborated to prototype two hybrid museum experiences. The museum in question, the Museum of Yugoslavia, was chosen as a particularly challenging environment, being home to a collection of artefacts related to Josip Broz Tito, the ruler of a country that no longer exists in a region that has seen many decades of political and often violent struggle. The two games that we report on were designed to bring in perspectives and reflections on the museum that were not strongly represented in the exhibition. As such, they enabled very different experiences compared to merely visiting the physical exhibition. While the games were designed and tested in collaboration with museum staff, we will show how the opportunities presented by hybrid technology also became challenges, from precisely this perspective. We will explore the following research question: \textit{What challenges may arise when designing hybrid museum experiences for critical play?}

\section{Related Work}
\subsection{New and Critical Museology}
In Peter Vergo’s introduction to the edited collection The New Museology, the author complains that “what is wrong with the `old’ museology is that it is too much about museum \textit{methods}, and too little about the purposes of museums”~\cite[~p. 3, emphasis in the original]{Vergo1989}. Vergo's interdisciplinary, multivocal, and multifaceted approach to a “new museology” emphasises the importance of values; societal assumptions, norms, and contexts; and stresses the importance of turning the one-way delivery of supposedly neutral information into a dialogue among the many stakeholders whose values, contexts, and stories have often been overlooked.

Today, the term “critical museology” ~\cite{Shelton2013} is often preferred. This perspective shares most if not all of these concerns. Anthony Shelton describes critical museology as divorced from the “operational museology”, the “methods” so well-rehearsed to Vergo’s ears, and concerned instead with increased accessibility, the representation of multiple groups, decolonisation, and the sharing of curatorial authority with visitors and the general public ~\cite{Shelton2013}. As part of this general turn towards a “new” or “critical” museology in the past 20 to 30 years, a central question emerged about the role of the \textit{object} in the museum – formerly unquestionably their primary focus – and the \textit{idea} (which may include not only the curator’s opinion, but  bringing together of a plurality of perspectives) ~\cite{weil_proper_1994, witcomb_side_1997}. In response, museum institutions have increasingly shifted their focus from highlighting physical collections to highlighting stories and experiences they can share with their audiences~\cite{hooper-greenhill_museums_2000}. This can include a higher emphasis on stimulating the senses; David Howes for example argues for a “sensory museology” in which visitors can interact with museums using all five basic senses, possibly facilitated by digital technologies~\cite{howes_introduction_2014}. Examples of this “sensory museology” include Andermann and Simine’s~\cite{andermann_introduction:_2012} work on memory-based museums and their focus on evoking emotions in their visitors. Both story-based approaches and sensory approaches share a fundamentally participatory ambition, in which the visitor is envisioned as possessing agency and scope for some form of interaction ~\cite{simon_participatory_2010}. This is especially interesting when the curated work in question is itself interactive (see e.g.~ \cite{giannachi_san_2009}). Some of the more wide-ranging practical implications of “new” or “critical” museologies have been studied within the museum literature, such as how a museum’s curatorial intent shapes visitor experience~\cite{tzortzi_movement_2014}, how people move through cultural heritage spaces in relation to other visitors~\cite{vom_lehn_configuring_2001}, and the ethical concerns that arise when visitors engage with a museum by creating their own digital content~\cite{kidd_space_2017}.  

\subsection{Digital Interactions in Museum Experiences}
The introduction of digital technology has proven to be relevant in addressing the concerns of these recent approaches to museology for a range of different reasons: some digital technologies can encourage participation and a sense of contribution, increase accessibility of certain collections to certain audiences who previously had no means of experiencing the museum or its objects for themselves, and strengthen the notion that the idea the object raises in the community is at least as worthy of engagement as the object itself. 

In HCI and interaction design, leading researchers Eva Hornecker and Luigina Ciolfi describe how museums have long been a "fertile research ground for Human-Computer Interaction research"~\cite[][p. xv]{hornecker_ciolfi_2019}. Some research has explored approaches to soliciting participation from museum professionals~\cite{avram_co-designing_2016, ciolfi_articulating_2016, risseeuw_authoring_2016}, but most have focused on visitor experiences. Interpersonal and social matters highlight the contextualised nature of museum visits, notably how families orient themselves to digital devices and respond to crowded points of interest~\cite{rennick-egglestone_families_2016}, how groups of visitors manage a coherent experience~\cite{tolmie_supporting_2014}, and how visitors choose what to photograph for posting on social media~\cite{hillman_situated_2015}. However, we see little effort within HCI to address broad, societal questions raised by museums showing work of a highly contested nature, except perhaps where "dissenters" such as Holocaust deniers would be roundly condemned by the wider population within which their museums are situated (e.g.~\cite{Ma2015}).

Early technological interventions often extended the interactivity of the traditional audio guide. However, the introduction of digital guide applications for visitors led to concern among some museum professionals, who worried that the technology would steal too much of the visitors' attention away from the museum exhibits. As early as 1996, Walter lamented the consequences of introducing "electronic guides": "Visitors became absorbed in their electronic guides, interacting less with their companions and less with the objects on display" ~\cite[][p. 241]{walter_museum_1996}. In subsequent years, similar concerns have often resurfaced, frequently referred to as “the heads-down phenomenon” \cite{hsi_study_2003, lyons_designing_2009, wessel_potentials_2007,petrelli_2013_integrating}. In recent years, some research has focused on the use of technology to facilitate group interactions in museums, helping users in their attentional "balancing act" \cite{woodruff_electronic_2001} between focusing on their phones and their surroundings \cite{fosh_supporting_2016,katifori_2020,eklund_shoe_2020}. 

\subsubsection{Hybrid Experiences}
We see promise in the area of HCI research of "mixed reality" or "hybrid" design, as discussed many years ago~\cite{bannon_hybrid_2005} and expanded upon in the context of mixed reality~\cite{Benford2011}. These researchers take advantage of the many affordances of digital technologies, particularly virtual and/or augmented reality, but design the experience in combination with the user's embodied presence in the particular physical space and social context of their surroundings. The literature includes notable examples of installation-based hybrid experiences in museums, from objects that beg passing visitors to be put on display~\cite{Marshall2015} to replica objects that guide visitors through the museum according to the perspective of their choice~\cite{Marshall2016} and objects situated on a context-aware backdrop that allow visitors to tell their own stories about them on augmented reality-enabled tablets~\cite{Darzentas2018}.

Increasingly, visitors’ own smart devices are being leveraged in mixed-reality design processes that actively seek to maximise a "heads-up" approach. Although there are still some significant socio-economic as well as simple preference-based barriers to full uptake~\cite{Petrelli2018}, digital interventions that take place on visitors’ own personal devices allow those visitors to co-create their own experiences using familiar tools from their own everyday lives~\cite{ciolfi_including_2008, magnenat-thalmann_virtual_2005}. One simple but striking example is an app that lets selected paintings overlay their style onto images of the everyday world outside or extend themselves into their virtual surroundings~\cite{Hurst2016}. Another used Artcodes~\cite{meese_codes_2013}, a QR-style code that can be designed in any artistic style, in one case to create unique interactive experiences for many hundreds of visitors to the Tate gallery in London~\cite{preston_enabling_2017}, and in another to link drawings visitors placed next to museum objects automatically to unique webpages holding their own spoken stories about those objects~\cite{Ali2018}. Another, which used photos and private audio messages between friends, allowed users to "gift" each other objects from the museum that they felt their receiver would like, often leading to a notable change in their behaviours and attitudes towards the museum experience
~\cite{spence_seeing_2019}. We sought to explore a similarly simple, private, smartphone-based experience, primarily because it would allow for the private expression of what might socially be perceived as unwelcome or even forbidden reactions to the strongly contested histories in our partner museum.

\subsubsection{Frictional Hybrid Experiences}
While most work on hybrid experience design focuses on the challenges related to creating as integrated experiences as possible \cite{benford_interaction_2009}, there is also a strand of interaction design research that has investigated ways in which the physical and digital aspects of an experience need not always be fully aligned. In early work on seamful design \cite{chalmers_2003_seamful,chalmers_2004_seamful}, Chalmers et al. argued that it may not always be beneficial to hide the inner workings of the digital infrastructure from users. Rostami et al \cite{rostami_2018_frictional} explored how performance artists will sometimes combine VR and real-world experiences in ways that capitalize on, rather than hide, the friction between the physical and the digital. Work on experience trajectory design places particular emphasis on how to design transitions between different media [7] to be visible and actionable, but also engaging. In the context of museum experiences, Fosh et al \cite{fosh_see_2013} suggested an approach to overlaying digital experience on sculptures in which visitors followed a five-stage journey through each exhibit – approach, engage, experience, disengage and reflect – with the official interpretation only being revealed during the reflect stage. These examples show that it is critical to understand how the digital and physical are connected in hybrid experiences, and that a fully integrated experience is not always possible or the best solution. 

\subsection{Games and play in museums}
There is a growing interest in games and play in museums \cite{beale_museums_2011,kidd_immersive_2018}. The idea to use play in museums is influenced by several different developments. One is the proliferation of learning theories that emphasize that children play to learn \cite{hein_learning_1998, nilsson_playing-exploring_2018, danniels_defining_2018}. Another development is the work towards opening up museums for co-creation and participation \cite{simon_participatory_2010}, which goes hand in hand with the process of digitalization and digitisation \cite{parry_recoding_2007} enabling new applications of play in museums. 

Over the years, a large number of serious games have been developed for museums, from simple web-based games to larger Virtual Reality productions\cite{anderson_developing_2010}. Mobile or pervasive museum games often tend to fall under the category of the scavenger hunt, where players follow clues and solve puzzles \cite{avouris_review_2012}. A popular example is The Murder at the Met Scavenger Hunt at the Metropolitan Museum of Art in New York \cite{kim_chapter_2015}. Experiments have also been done with role-playing games \cite{paay_location-based_2008}, storytelling games \cite{vayanou_how_2019}, as well as games fostering different variations of playful and open-ended interactions with museum artefacts \cite{wakkary_situated_2007,coenen_museus:_2013}. The British artist group Blast Theory has explored the intersection of art and pervasive games in a variety of museum and heritage settings through works such as Ghostwriter \cite{blast_theory_ghostwriter_2011} and Fixing Point \cite{blast_theory_fixing_2011}.

However, the introduction of games into a museum environment is not without its challenges. Bergström et al. \cite{bergstrom_gaming_2014} discuss the challenges of using gamification to facilitate informal learning, while preventing the game from distracting from the learning. Wakkary and Hatala \cite{wakkary_situated_2007} equally point out that it is important that the playfulness induced by the design is not perceived to be separate from the museum environment to the point where it is distracting or does not make sense for the players. However, as we saw from the examples of so-called "frictional hybrid experiences", there are situations where is makes sense to use play deliberately to create tensions and induce a critical awareness, rather than to align it completely with the physical environment. One conceptualization of such play is what we would like to present next, namely \textit{critical play}.  

\subsection{Critical Play}
Critical play is a form of play in which norms and conventions are deliberately being challenged. Games scholar Mary Flanagan \cite{flanagan_critical_2009} has forwarded it as a broad concept which encompasses a wide range of play activities  from artistic play practices – such as techniques used by the Surrealists – to engagement in modern videogames designed for political, aesthetic, and social critique.

The study of critical play can roughly be divided into three parts: Firstly, the study of \textit{transgressive play}, which is concerned with play against the `ideal’ or `implied’ player of a game, of bending rules and playing a game in ways not anticipated by design \cite{jorgensen_transgression_2019, sunden_play_2009, aarseth_i_2007}. When it is done deliberately as part of a critical political agenda, Dyer-Witheford and de Peuter name this activity “counterplay” \cite{dyer-witheford_games_2009}. Secondly, we have the study of \textit{critical games}, which are games that pose an alternative to mainstream games \cite{lindsay_critical_2014,bogost_persuasive_2007,frasca2007play}. Examples of these games include \textit{Phone Story} \cite{molleindustria_phone_2011} and other games by Molleindustria \cite{molleindustria_molleindustria_nodate}, as well as work by Anna Anthropy \cite{anthropy_witchio_nodate}, along with other games belonging to queer games studies \cite{ruberg_queer_2017}. Lastly, and most relevant for our study, is the study of \textit{brink play} \cite{poremba_critical_2007}, also named \textit{boundary play} \cite{nippert-eng_boundary_2005}. The critical potential in this form of play, lies in its power to let players explore personal, social, and cultural boundaries protected by the mindset and the social contract of "this is just a game". 

Traditionally, play is seen as an activity which takes place separately from the 'real' world. Renowned play scholar and historian Johan Huizinga describes play as creating "temporary worlds within the ordinary world, dedicated to the performance of an act apart" \cite[p. 10]{huizinga_homo_1998}. These worlds "within which special rules obtain"(ibid), are sometimes referred to as "the magic circle of play" \cite{stenros_defence_2012}. According to games scholar Cindy Poremba, by pushing or breaching the bounds of the magic circle, brink games not only force the awareness of explicit and implicit game rules, but of implicit and explicit non-game rules as well \cite{poremba_critical_2007}. This includes social norms and cultural conventions of the environment in which the play is situated. In this capacity, critical play gives players the opportunity to examine the cultural regimes in which they live \cite{sutton-smith_evolving_1999} and allows them to engage in small acts of deviance \cite{henricks_orderly_2009}. 

Transgressing norms in a playful way is central to comedy \cite{critchley_humour_2011} and to expressions of the carnivalesque \cite{bakhtin_problems_2008, bakhtin_rabelais_1984} but can equally be done in a more introspective and intimate manner \cite{ryding_silent_2020, ryding_play_2020}. Therefore, games that are designed to foster brink or boundary play employ a number of different approaches that range from the humorous and satirical, to the poetic and the more openly confrontational. This can be seen in the work by artists and designers such as Joseph DeLappe \cite{delappe_joseph_nodate}, Cory Arcangel \cite{arcangel_cory_nodate} and Brody Condon \cite{condon_brody_nodate} among others, as well as in numerous work from the Nordic Larp scene \cite{stenros_nordic_2010}.  

An example of a game which specifically explores critical play in a museum setting, is Art Heist. This is an interactive narrative piece that was developed at The New Art Gallery Walsall in October 2010. In this role-playing game the participants plan and carry out an art theft at the gallery: “Art Heist uses the transgressive qualities of art theft and the excitement of breaking into a gallery while posing its audience big questions about art: who is it for, who decides what’s good, why is it valuable and does the value of the art lie in the idea or in the object?” \cite{mees_art_2011}.

Art Heist, in its rather radical use of theft as a metaphor, points to how \textit{appropriation} is a crucial part of art practice as well as play \cite{henricks_orderly_2009, sicart_play_2014}. Interestingly, in a cultural heritage context, the term "appropriation" typically refers to borrowing or even stealing another culture’s artefacts and histories without permission, a deeply problematic challenge in the era of post-colonialism \cite{ziff_borrowed_1997}. In HCI, on the other hand, appropriation is a term with positive connotations, referring to how users may come to appropriate technologies for their own purposes during the course of practice \cite{dourish_where_2001, dix_designing_2007, carroll_identity_2001}. In Art Heist, participants needed to make their own decisions and actively use the space of the museum for their own objectives, something which is a typical characteristics for play \cite{henricks_play_2015}. In this sense, critical play opens up for new ways to interact with museums and their content, ways that can prove to be ethically challenging as historical contexts and artistic intentions are running the risk of being ignored. On the other hand, critical play can potentially be a powerful way for visitors to engage with contested heritage by fostering an active questioning of historical discourse as well as traditional museum conventions. 

Another notable example of particular relevance to the prototypes presented in this article, is the "Bad News Game" \cite{roozenbeek_fake_2019}, which invites players to take on the role of a fake news creator in order to learn about the techniques used in fake news. The creators of the game describe this as an "inoculation" strategy, positing that "preemptively exposing, warning, and familiarising people with the strategies used in the production of fake news helps confer cognitive immunity when exposed to real misinformation" \cite[p. 1]{roozenbeek_fake_2019}. Brenda Brathwaite's game \textit{Train} explores the phenomenon of complicity by tasking players with loading toy people onto train cars, only later to reveal that these trains are headed for Auschwitz \cite{brathwaite_mechanic_2010}. In a similar (albeit less extreme) vein, the game \textit{Papers, please} by Lucas Pope \cite{pope_papers_2013} puts players in the role of border bureaucrats in a fictional Eastern European country, in what Sicart calls "an exploration of totalitarian bureaucratic systems and the banality of evil" \cite[p. 151]{sicart_papers_2019}.

\subsection{Contested heritage}
This article focuses on the challenge of facilitating critical play with an extraordinary example of contested heritage: the The Museum of Yugoslavia, which is dominated by the objects, stories, and mausoleum of its controversial leader, Josep Broz Tito. Tito ruled Yugoslavia from 1953 until his death in 1980. He personally led the forces that defeated the Nazi invasion of his homeland. Yet he used guerilla tactics that could today see him condemned as a "terrorist". As a Communist, he aligned Yugoslavia with the Soviet Bloc, but alone amongst the Soviet satellite rulers, he allowed some freedom of movement and adopted some Western economic practices. Yet (again) he exerted his powers through a "cult of personality" culminating in a "Presidency for Life", leaving little room for dissent. After his death, the country was ill-equipped to chart a peaceful course, culminating in the Bosnian War of 1992-1995. Domestically and internationally, Tito still polarises opinions. Some see him as a heroic force for good, while others see him as a brutal megalomaniac.

Within heritage studies, much work has explored the challenges with preserving and exhibiting "difficult heritage"~\cite{macdonald_difcult_2008,drozdzewski_memory_2016,jong_witness_2018}. Viewed from the perspective of discourse theory, the concept of heritage itself may be considered as a contest among hegemonic discourses \cite{smith_uses_2006}. Silverman discusses the consequences of "socially engaged, politically aware study of the past that regards heritage as contested (and) recognizes the role of power in the construction of history (...)" \cite{silverman_social_2010}. Silverman sees cultural heritage as full of conflict and power struggles between different communities of practice, and as an ongoing process in which history is not just accounted for but also created as a political project. Studies of contested heritage have focused on studying the sites in which there are multiple actors involved and the material objects themselves can be of controversial character. In that sense contested heritage can relate to topics such as post-colonial ownership of artefacts \cite{blakey_new_1998, reid_whose_2002}, deciding what is and is not valuable to preserve \cite{silverman_cultural_2007}, and how to treat cultural heritage in light of tourism and increased public interest \cite{landry_touring_2011}.

Galaty~\cite{galaty_blood_2011} looks at how different stakeholders in sites of conflict, such as the Balkans, hold multiple interpretations of their cultural heritage and how important it is to make space for expression of those multiple points of view. Andrea Witcomb \cite{witcomb_understanding_2013} argues for the use of affective strategies to enable a form of historical consciousness that leads to a critical engagement with "difficult" topics such as histories of war and oppression. She promotes sensorial, embodied museum experiences that may trigger emotional responses, rather than exhibitions that use explicitly rational, information-based content on linear display. She underlines how the poetic production of unsettling experiences requires visitors to engage both intellectually \textit{and} emotionally, potentially leading to deeper and more meaningful experiences for the visitors. 

\section{Method}
Our approach has been one of ‘Research Through Design’ \cite{zimmerman_research_2007}, in which research findings emerge from design practice. We worked in a practice-led manner through the design, prototyping and review of two contrasting visiting experiences as a way of engaging with the changing face of the modern museum and the role that hybrid technologies might play in reshaping the visiting experience. These designs were developed in the form of smartphone apps: \textit{Twitto} and \textit{Monuments for a Departed Future} - both of which were designed for and tested at the Museum of Yugoslavia. 

Our design team included a professional game design company as well as academic researchers with design, human-computer interaction (HCI) and museology backgrounds. We worked in situ so that physical elements of the interventions were visible to regular users. However, because these were pilot studies regarding an extremely sensitive and contested history, they were not released to the general public. Even so, we consider them to be inspired by the aims of research "in the wild"~\cite{benford_performance-led_2013, crabtree_introduction_2013} to the extent that we worked with a museum that would help us engage with its current challenges of contemporary museum interpretation. We worked in its space and with its staff and visitors over the course of a year to inform, test and challenge our designs. 

The two designs that we present are suggested not as finalised design solutions, but rather as experimental prototypes used to critically probe the issues of critical play in a complex and challenging setting. In describing the prototypes, we present both the design challenges that we wrestled with as a team and the feedback that we gathered from curators and visitors. Due to the sensitive nature of the heritage within which we were working, we tested our prototypes on invited participants, all of whom could provide expert insight from the perspectives of museum professionals, designers, and students. Therefore, while we present some of their feedback and insights gained from it, we do not intend this as a substitute for evaluation by members of the general public.

Participants were interviewed after the test sessions, following a semi-structured interview guide. Interviews were audio recorded, transcribed and analysed by members of the research team to identify themes for the analysis. In some of the tests researchers were also observing the participants, and insights from these observations are included in the analysis where relevant.

The designs presented here formed part of a larger, cross-disciplinary research project funded by Horizon 2020, exploring hybrid museum experiences, called the GIFT project \cite{back_gift:_2018,lovlie_gift_2019}. The project included extensive collaboration between researchers from three universities, museum professionals from a wide range of European museums, as well as two design companies. In order to anchor the research and design in the concrete needs of museums, the project set up an action research process with 10 museums in Europe and the US. Each of the two design companies partnered with one of the participating museums to develop experimental prototypes intended to explore innovative formats for hybrid museum experiences. The first of these sub-projects focused on the concept of hybrid gifts, and resulted in the \textit{Gift} app developed by the UK company Blast Theory, presented in \cite{spence_seeing_2019}. The second sub-project, led by the Serbian design agency NextGame – which resulted in the \textit{Twitto} prototype discussed in this article – focused on "playful appropriation". This concept was explored in parallel by one of the authors - Karin Ryding - as part of her PhD research, resulting in the \textit{Monuments for a Departed Future} prototype, first presented in \cite{ryding_monuments_2018}. 

\section{The Museum of Yugoslavia}
The Museum of Yugoslavia, situated in the Serbian capital of Belgrade, has more than 100,000 visitors a year and is the most visited museum in Serbia. The museum is located on the grounds of the former communist leader Josip Broz Tito’s palace, and houses the grave of Tito and his wife. From our initial visits at the museum, an impression emerged that the dominating presence of Tito’s legacy presented a challenge for the museum curators (see "Contested Heritage" above) in developing the more critical perspective they aim to present. As stated in the museum's strategy documents:

\begin{quote}
Principles, values, interpretation and even heritage itself are changeable categories which are created in relation to the contemporary context — ideology, politics, the economy and scientific models. Because of this, we encourage CRITICAL THINKING and presentation of diverse views of the same events, documents or data. \cite[][(caps in the original)]{noauthor_strategic_2014}
\end{quote}

The museum representatives in the project have expressed a desire for the museum to reinvent itself as a modern institution that explores how digital technologies may help enable a more participatory and dialogue-based visitor experience. The tensions of such a vivid and recent history of contested heritage presented us with rich opportunities for using hybrid experiences to tell alternative stories and present alternative experiences.

\section{Twitto}

If the Yugoslavian communist leader Tito had Twitter, how would he have used it? Would he have been as sophisticated using social media as he was in the propaganda techniques of his own time? This is the playful premise of \textit{Twitto}, a game that invites visitors to put themselves in the shoes of an autocratic dictator and learn about propaganda through building their very own cult of personality. Exploring the historical exhibitions of the Museum of Yugoslavia, visitors are invited to create their own manifestos, posters, autobiographies and other items of propaganda for whichever cause they choose to make their own – be it "Death to capitalism", "girl power" or "pineapple on pizzas". In the following we will present the design of this game, through the design process and rationale, the final prototype and the evaluation.

\subsection{Design process and rationale}

It is characteristic of the design process of \textit{Twitto}, as for the rest of the GIFT project, that the process involved a broad cross-disciplinary team of designers, developers and university researchers, but also involving extensive collaboration with museum representatives. The design team initiated the process by inviting the university researchers and representatives of the museum to a workshop, in order to explore the museum's needs and the opportunities and challenges posed by the project. In the following 3 months, the company worked iteratively in collaboration with some of the university researchers to develop a first prototype, which was tested twice with museum representatives - first as a paper prototype, later as a fully functional app in a workshop at the museum on 28-29 June 2017. Experiences from this test led to a complete redesign of the concept, leading to a new prototype that was tested with a group of invited test participants on 22 February 2018.

From the start of the process, the museum representatives made it clear that they were not interested in bringing in play or technology just to attract new visitor groups. Rather, their motivation was to find ways for visitors to engage more deeply with the museum, seemingly reflecting the "forum" ideal of new museology, in which the museum should see its mission as facilitating dialogue and debate. In the first workshop one of the museum representatives put it as follows:

\begin{quote}more people is not basic goal. Definitely not. It’s to get those values that we don’t have right now. Maybe you will hear stories that you don’t know right now. Maybe you see perspectives you didn’t see until this point. (...) We are trying to unlock one more perspective, one more meaning, what this object means to another person. This is how we see technology as a tool.
\end{quote}

Throughout the design process, three questions in particular stood out. First, how would the digital experience be linked to the physical exhibits, \textit{technically}? Second, how would the digital interaction (the game) be connected with the physical exhibits, \textit{conceptually}? Third, how would the game facilitate play - and in particular, \textit{critical play}?

Regarding the first question, the design team decided early on to use the Artcodes platform \cite{meese_codes_2013} as a prototyping tool, as this supports the design of scannable markers that can be shaped flexibly to match the desired aesthetic, and can be redefined dynamically to allow visitors to reshape the interactions enabled by the marker. The team experimented with a number of different ways to integrate the Artcode markers into the experience, as explained below.

\begin{figure}
  \centering
  \includegraphics[width=0.5\linewidth]{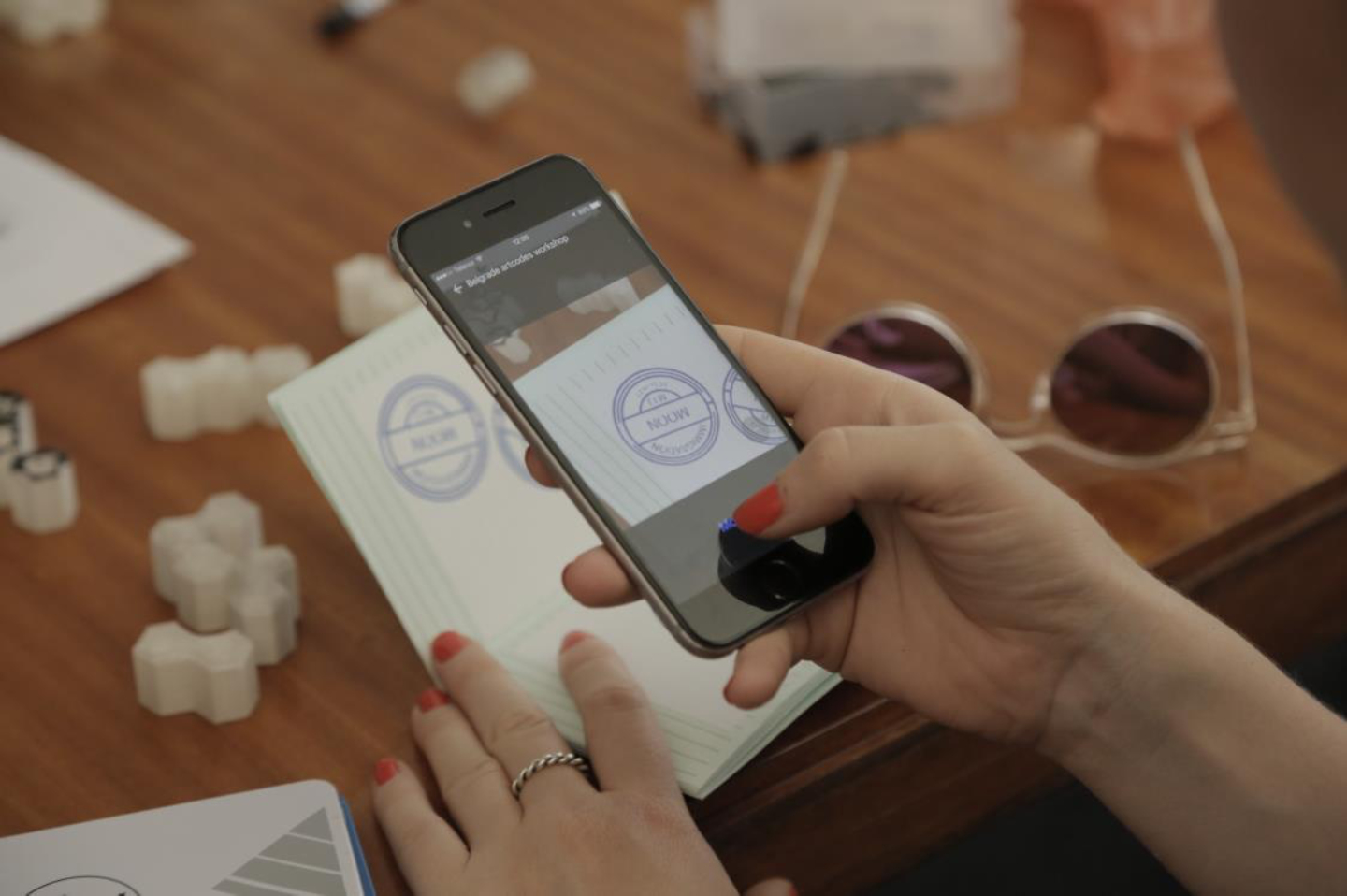}
  \caption{Scanning a passport stamp in the first version of \textit{Twitto}.}
  \label{passport}
\end{figure}

This work happened in close connection with the second question, about the conceptual connection. Initially, the team focused on one prominent feature of the museum: Its large collection of rare and valuable objects given to Tito as gifts from foreign leaders and diplomats, exhibited alongside (and often overshadowing) the museum's presentation of key moments in Yugoslavian history, through the two World Wars, a communist dictatorship and finally the country's disintegration in civil war. This inspired the first prototype, which used physical (but fake) passports and ink based stamps as the central technique and metaphor. During the cold war, Yugoslavian passports were known for being among the few passports of the world that would allow free travel both in the east and the west. In the exhibition, a number of passport stamps were placed next to gifts from leaders of respective country – e.g. a passport stamp representing the USA next to a gift from the US president, etc. Before entering the exhibition, visitors were handed a physical (fake) passport, and instructed that the goal of the game was to try to find as many physical stamps in the exhibit as possible and use them to stamp their passport (see Figure \ref{passport}). The stamps produced marks that were Artcodes, which could then be scanned using the game app, leading the visitor to a historical photograph and a short text that presented a short exchange between Tito and the foreign leader. In presenting these exchanges, the designers were attempting to explore Tito's skilled use of propaganda and relate it to contemporary political rhetoric. Thus, the narrative was based on the idea that if Tito had been alive today, he would probably have been a skilled user of social media such as Twitter - hence the name of the app, Twitto. Each scenario ended with a prompt, inviting the visitor to put themselves in the role of Tito, tweeting about this incident – what would Tito tweet?

\begin{figure}
  \centering
  \includegraphics[width=0.5\linewidth]{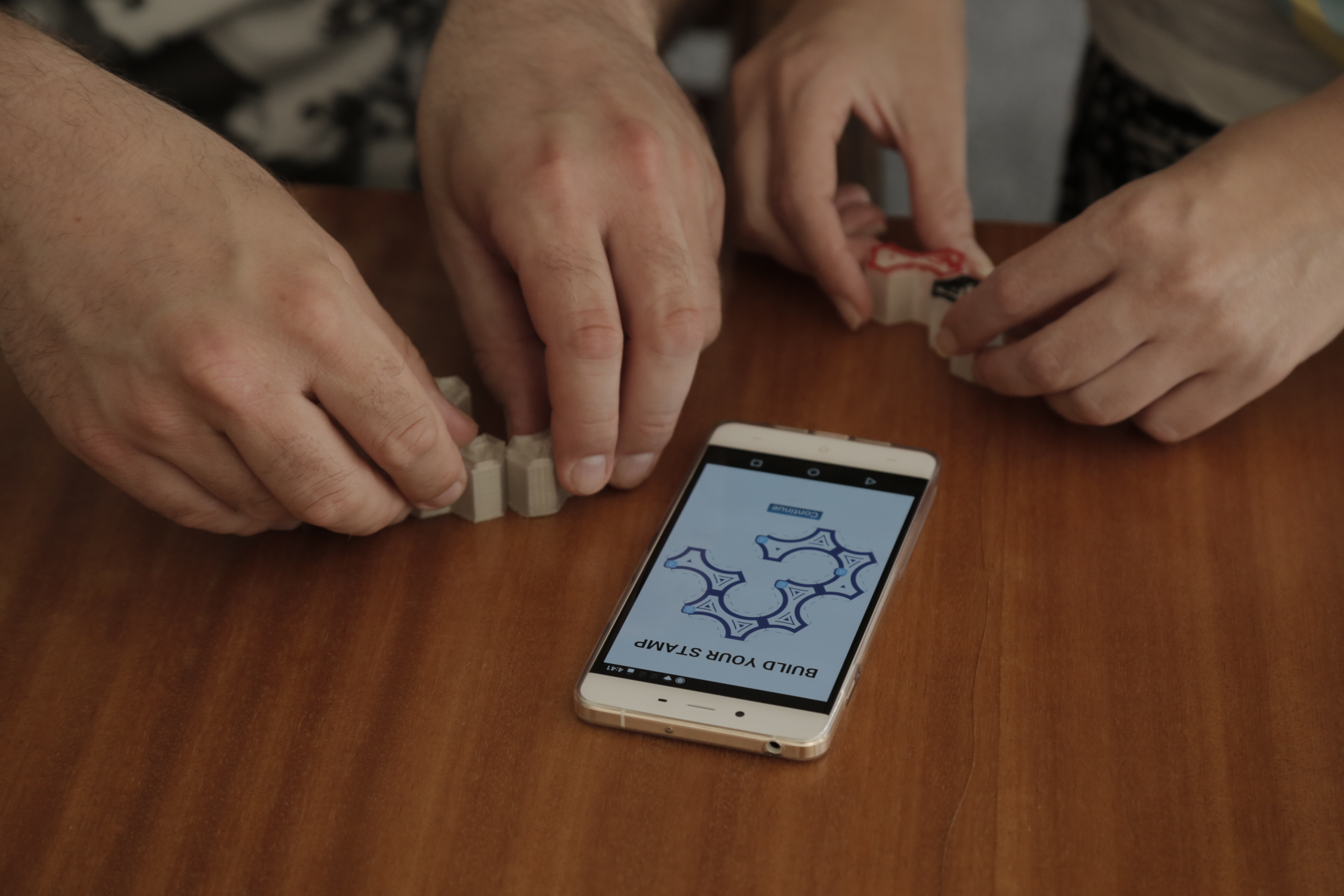}
  \caption{Assembling an Artcode stamp out of small building blocks.}
  \label{stamps}
\end{figure}

This prototype was tested in the workshop 28-29 June 2017, by participating researchers as well as museum representatives. While all participants were enthusiastic about the use of passports as both conceptual and technical link between the physical and digital, it turned out to be problematic in practice as the stamps often would not be precise enough to be scanned by the app. Another technique was tested out at the same workshop, consisting of small "building blocks" that could be assembled by a visitor to create their own unique Artcode stamp (see Figure \ref{stamps}). These could be used to leave a stamp on blank posters hung in the exhibition – and visitors could then post a digital message using the Artcode app, that other visitors would see when scanning their stamp. However, this format also turned out to be difficult to use in practice. Museum representatives made it clear that it would be impossible for them to allow visitors to use ink stamps in the exhibition, due to the risk to the precious objects on display. Furthermore, the team wished to facilitate a deeper critical engagement with the historical content of the exhibition.

Based on these experiences, a new prototype was designed and implemented. In this version, the technical linking between physical exhibits and digital content was simplified: instead of using stamps, the design team created stickers that could be placed next to exhibits relevant to the game and scanned to trigger interactions (see Figure \ref{wanted}). Furthermore, the new prototype included a stronger focus on Tito's use of propaganda, in particular his "cult of personality". The app presented Tito's history, from early childhood through his days as a rebel leader to his long rule as communist statesman, through a series of "chapters". The app would teach visitors about Tito's propaganda tactics and then challenge them to put themselves in the role of an authoritarian leader constructing their own propaganda messages.

\subsection{Twitto: Final prototype}
\begin{figure}[h]
  \centering
  \includegraphics[width=0.5\linewidth]{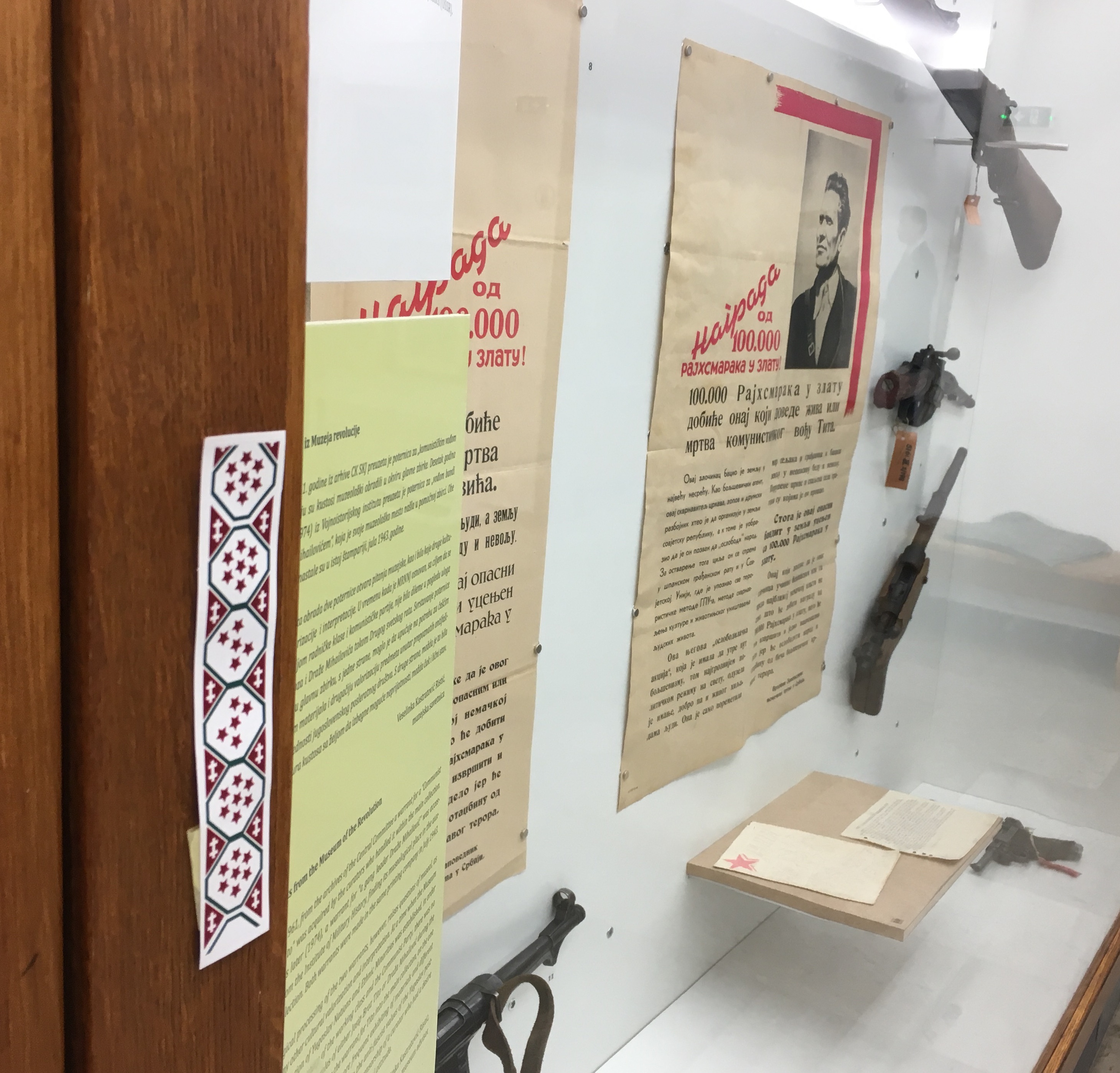}
  \caption{Artcode from Twitto posted next to a famous "Wanted" poster from WWII, relating to Tito's time as partisan leader fighting the Nazi occupation.}
  \Description{Artcode from Twitto posted next to a famous "Wanted" poster from WWII.}
  \label{wanted}
\end{figure}

\begin{figure}[h]
  \centering
  \includegraphics[width=0.2\linewidth]{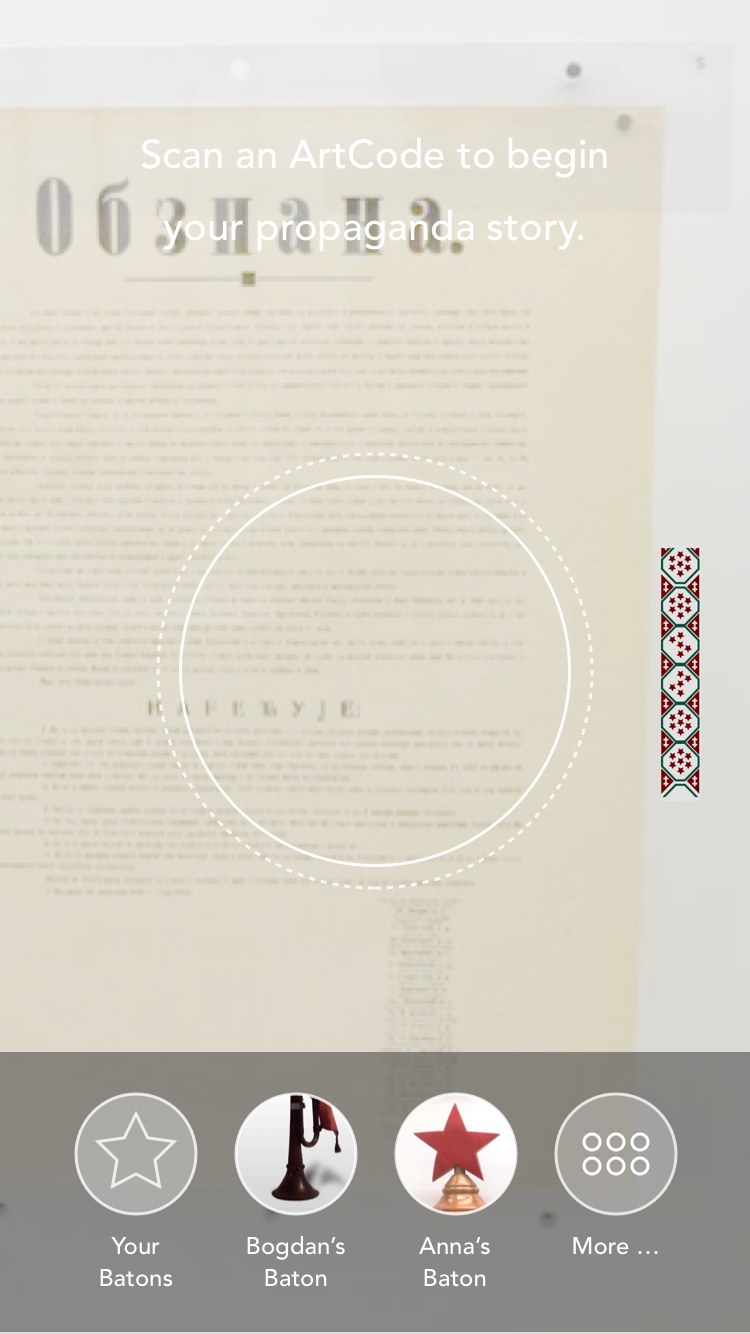}
  \includegraphics[width=0.2\linewidth]{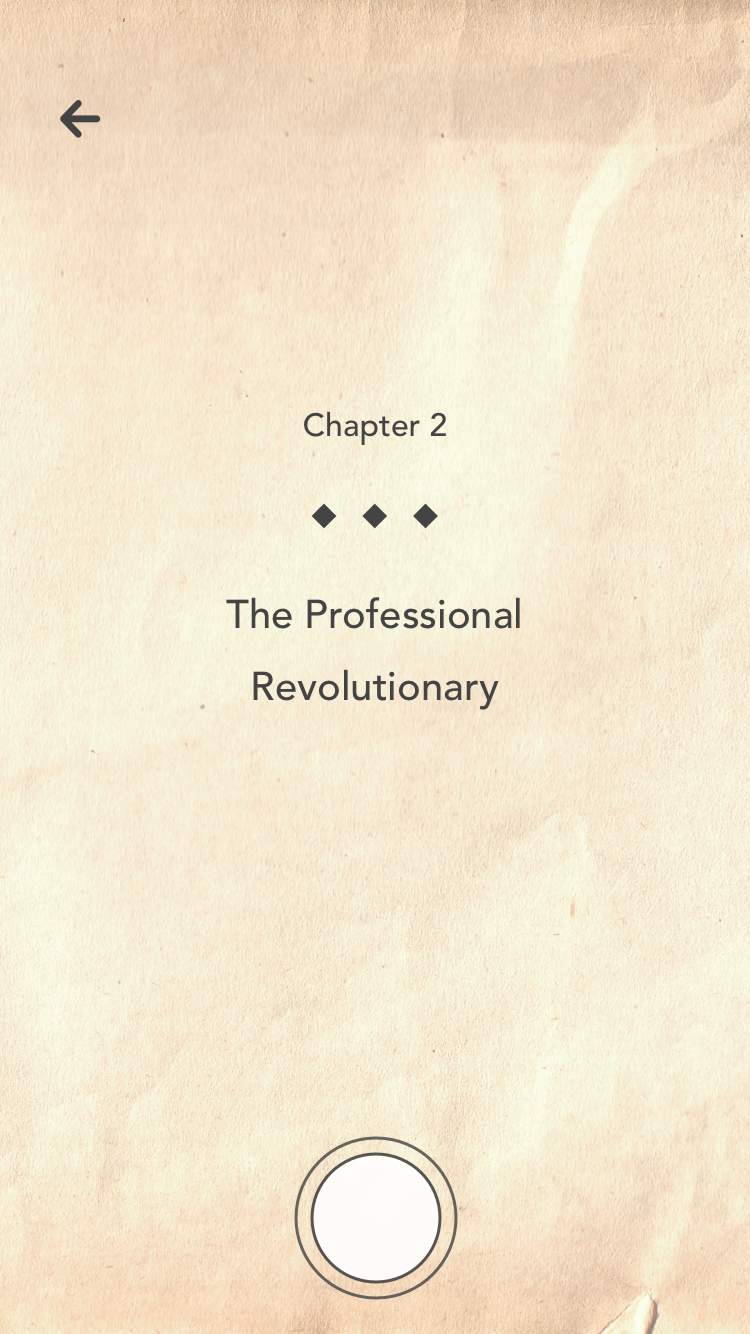}
  \includegraphics[width=0.2\linewidth]{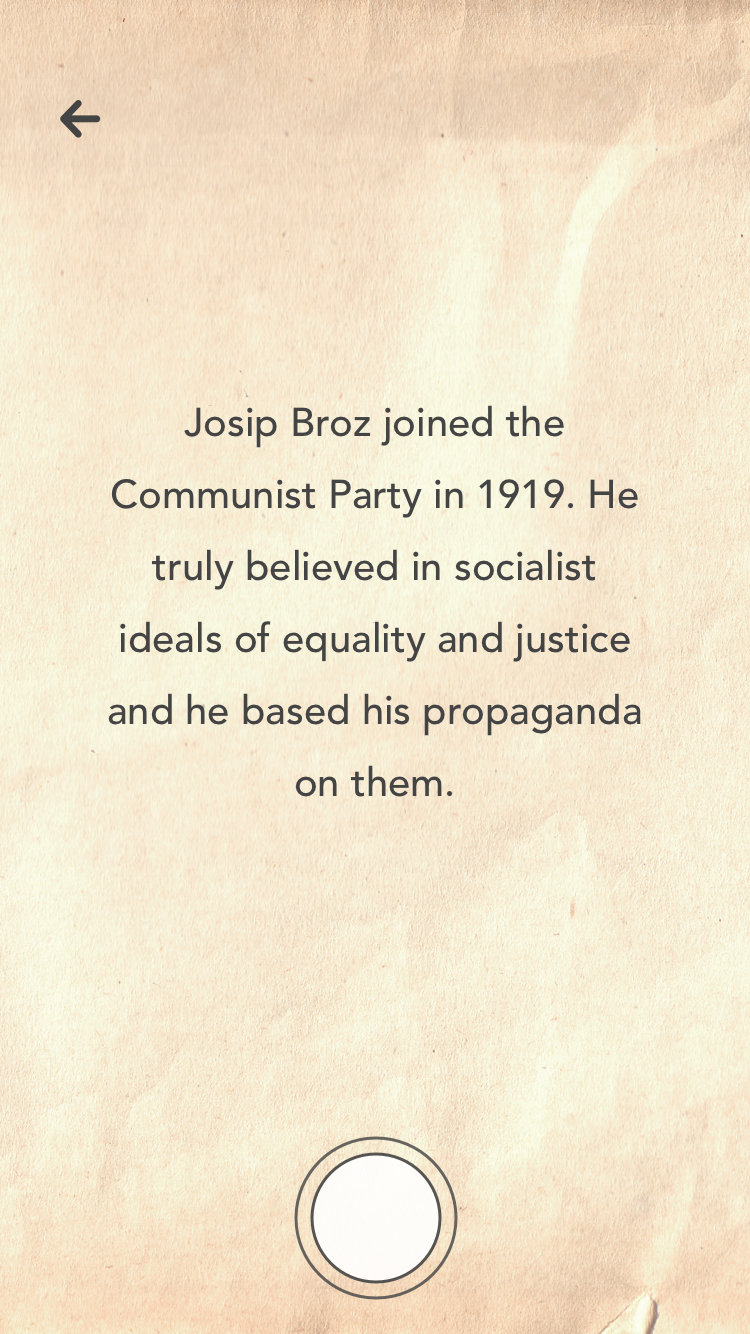}
  \includegraphics[width=0.2\linewidth]{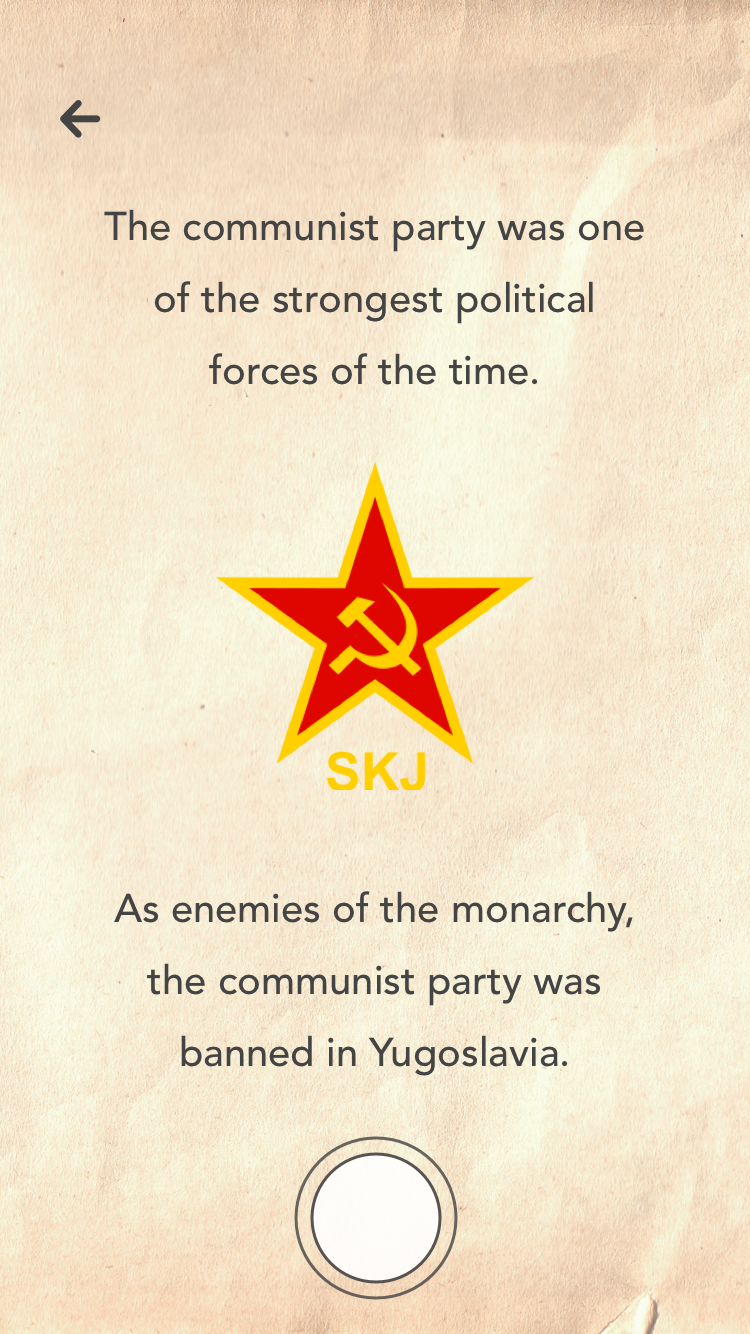}
  \includegraphics[width=0.2\linewidth]{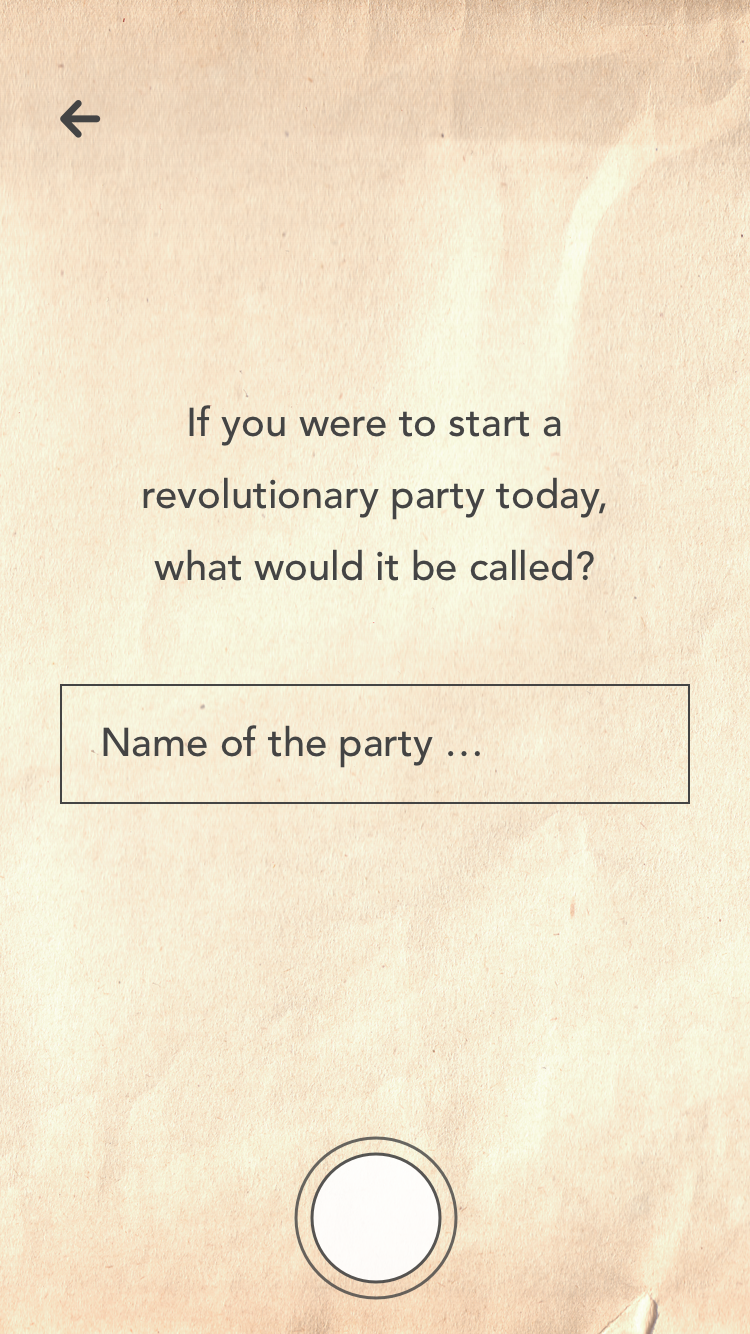}
  \includegraphics[width=0.2\linewidth]{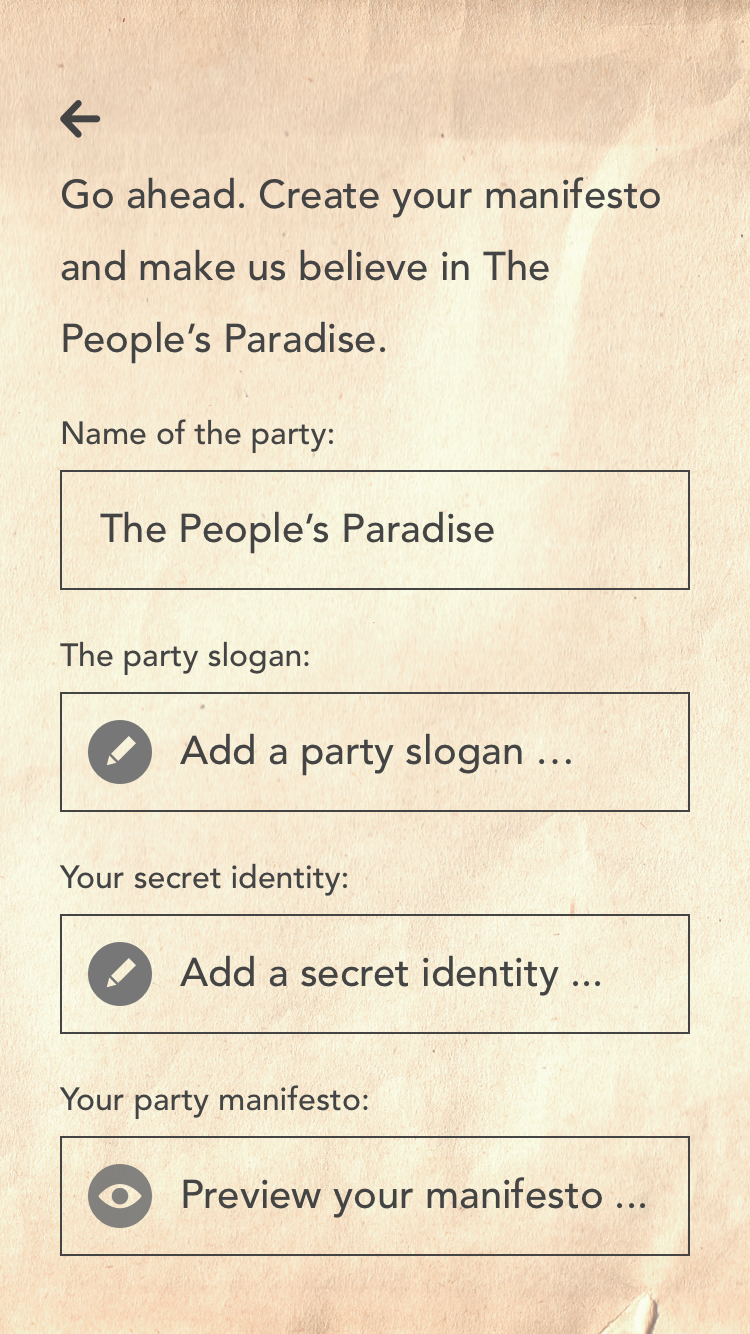}
  \includegraphics[width=0.2\linewidth]{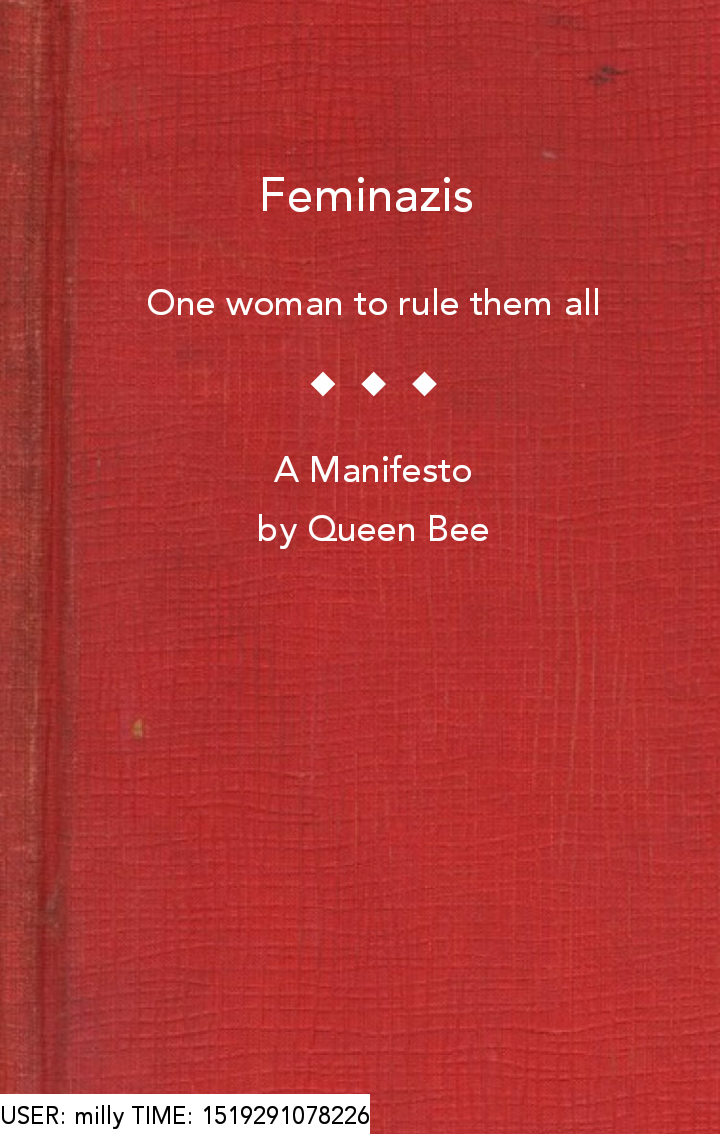}
  \caption{Screenshot from the Twitto app. When an Artcode is scanned, the app opens a series of screens that present a "chapter" in Tito’s life, and prompt the player to create their own propaganda item – in this case, a manifesto. The last image is a manifesto created by one of the test players.}
  \Description{Screenshot from the Twitto app.}
  \label{twitto_screenshots}
\end{figure}

\begin{figure}[h]
  \centering
  \includegraphics[width=0.2\linewidth]{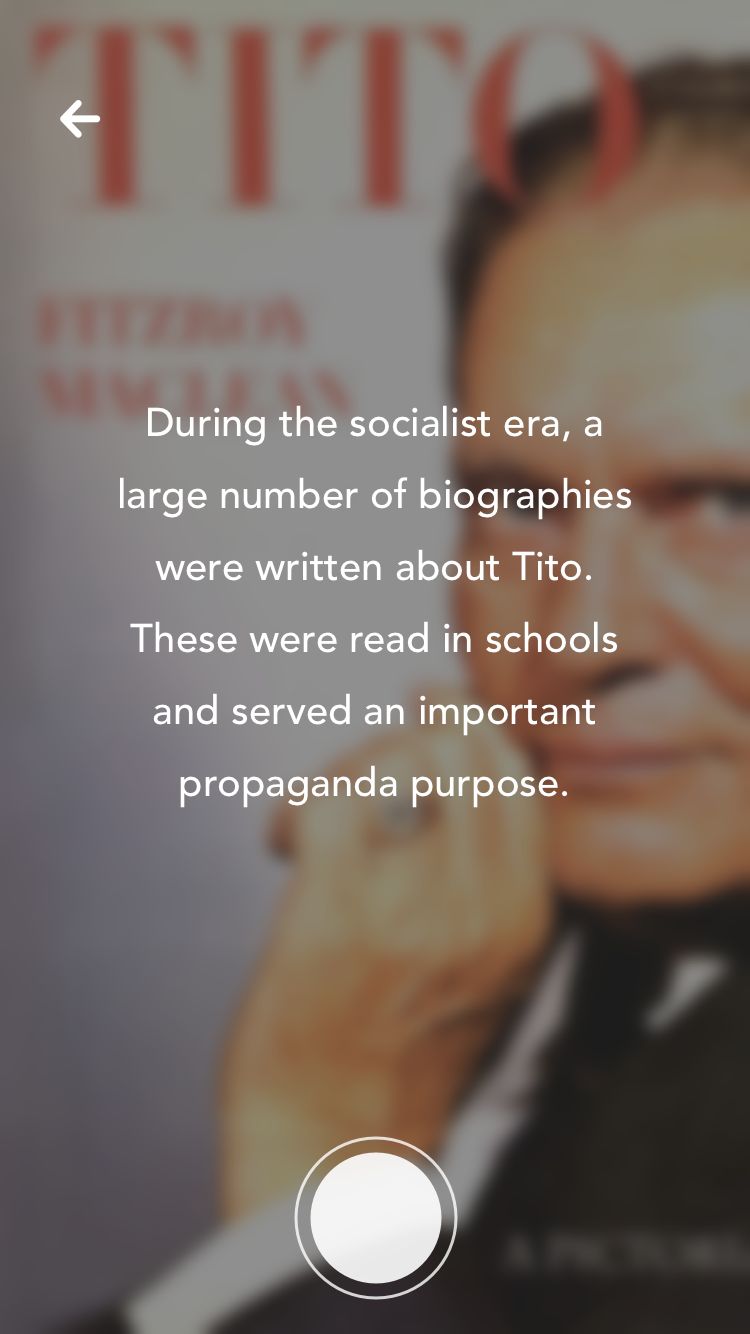}
  \includegraphics[width=0.2\linewidth]{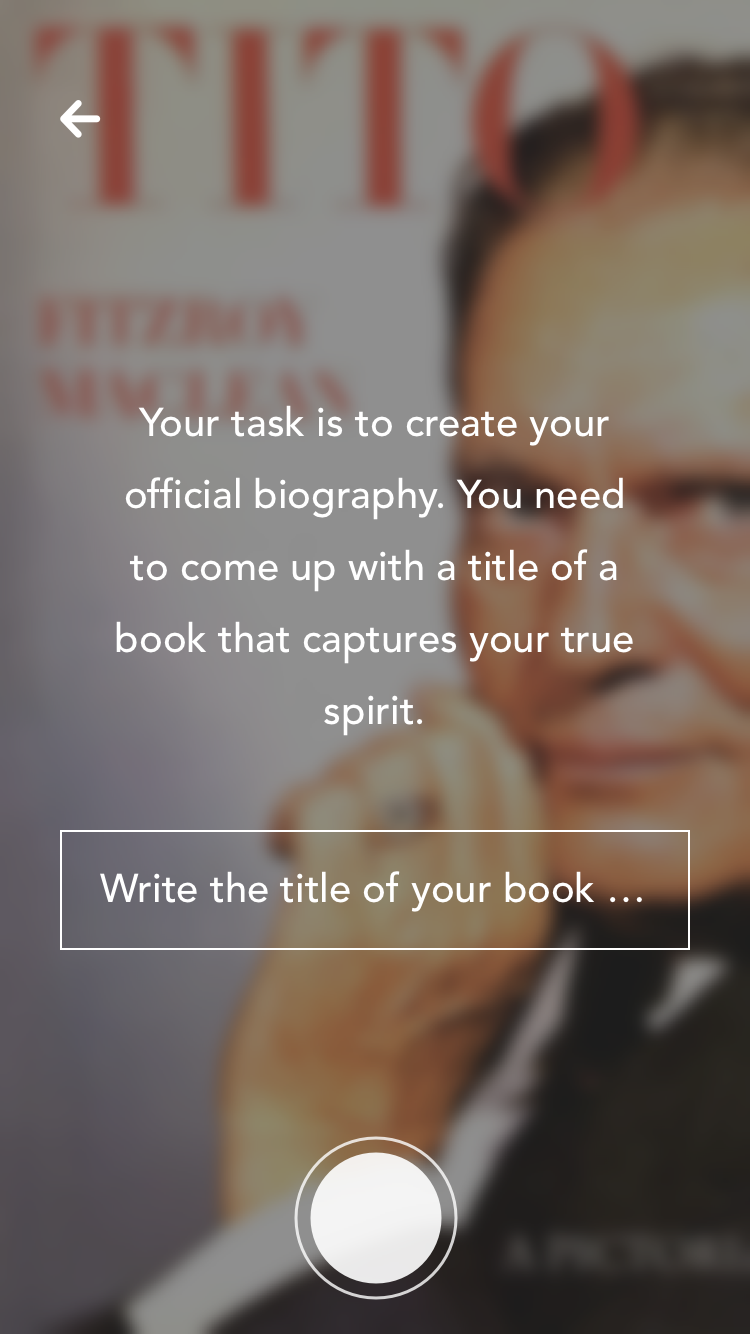}
  \includegraphics[width=0.2\linewidth]{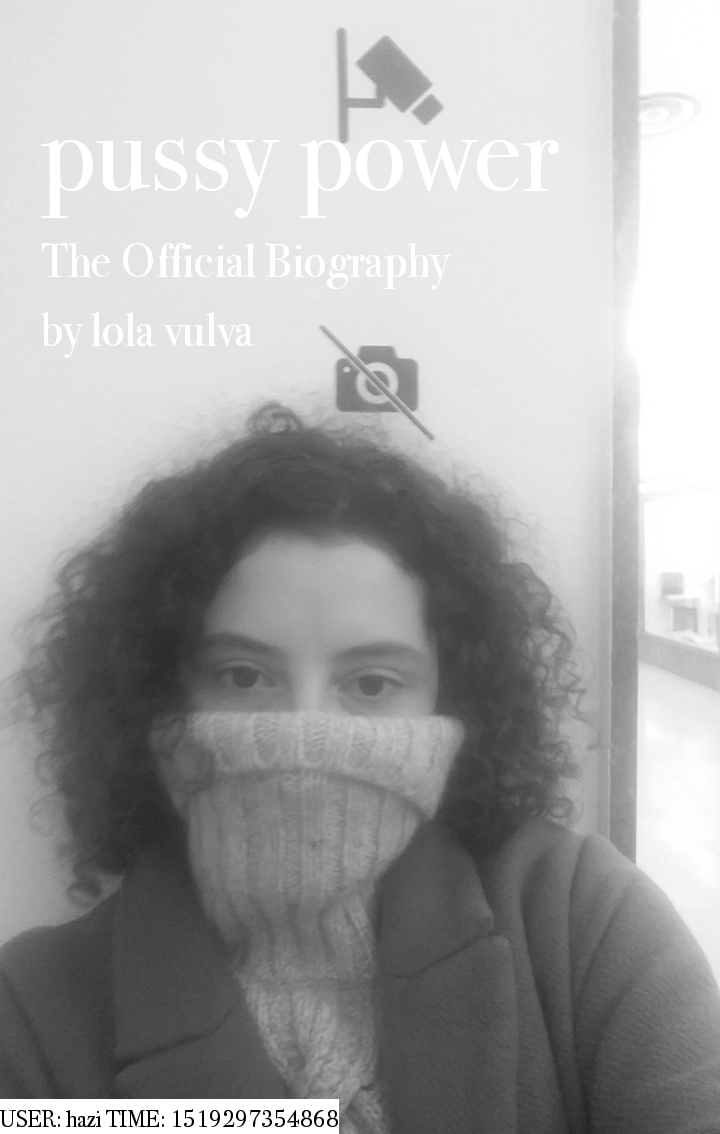}
  \caption{A challenge from the Twitto app, together with the response from one of the participants.}
  \Description{A challenge from the Twitto app, together with the response from one of the participants.}
  \label{prompts}
\end{figure}

Below, we briefly describe the prototype as tested in February 2018. Twitto can be described as a single-player role-playing game, in which the player is cast in the role of a resistance leader and eventually dictator. Going through the museum, the player may scan stickers with Artcodes posted next to artefacts of particular significance in the story (or myth) of Tito (see Figure \ref{wanted}). The Artcodes were designed to resemble insignia on partisan uniforms from Tito's rebel army during World War II. Scanning a sticker would open a new `chapter' in the game, starting with a few short sentences presenting one period of time in Tito’s life and the significance of the scanned artefact (Figure \ref{twitto_screenshots}). After this the player is prompted to put themselves in Tito’s shoes, e.g.: If you were a political resistance leader, what would your party be called? What would your propaganda poster look like? For each "chapter" in Tito’s biography, the player is tasked with answering a series of prompts as in Figure \ref{prompts}. The input is fitted into a predesigned template, resulting in the player assembling a propaganda item – a poster, party manifesto, book cover, etc. Thus, the game offered a series of creative challenges to the player, framed to fit into a propaganda format that was intended to facilitate both light-hearted play as well as critical reflection on the nature of propaganda both in the history of communist-era Yugoslavia and today.

\subsection{Evaluation}
In order to test the feasibility and acceptability of this prototype, Twitto was tested by a small group of 24 participants at the museum on 22 February 2018. Due to the nature of the prototype – dealing with sensitive topics, and being perceived as potentially challenging to the museum – we recruited test participants with relevant backgrounds from the networks of the design company and the museum representatives, and consisted mainly of students and professionals from the museum and design sector. Among the test participants 16 were female, 8 male. The test participants were invited to explore the exhibition on their own, using the app on test devices they borrowed from the researchers. After the test, the participants were debriefed in group interviews with researchers.

Twitto received much positive feedback from participants in the playtest, being described as ‘fun’, ‘creative’, ‘really cool’, ‘really funny’, ‘inviting’, ‘imaginative’, and ‘an unexpected experience’. Looking into the co-created content, we see that participants responded to the challenge of making propaganda for themselves by connecting to their personal lives at varying levels of seriousness. Some adopted a playful tone, others aimed for a more serious approach. For instance, when asked to give a title to a fictive manifesto, participants suggested titles like: ‘Revolutionary Cats: Cats are life’, ‘Gluten: Free gluten to everyone’ and ‘Justice: Death to capitalism’. 

Given the historic tensions and still simmering conflicts in several parts of the former Yugoslavia, the design team was concerned about causing offence when inviting visitors to the museum – and Tito’s grave – to engage in playful behavior that could include transgressions such as mockery and incivility. At the same time, the designers were equally worried about adopting an uncritically laudatory tone and therefore adopt the version of Tito’s history that had been used in propaganda. To our surprise, introducing playfulness to the exhibit did not cause offence; it was received with enthusiasm and gave plenty of food for thought in the follow up discussions with our test participants. However, in spite of these positive comments, the overall reception among the participants was negative. The main argument raised against the app was that it seemed to direct the players' attention away from the artefacts on display. "It was fun, it was inviting, but I expected it to be more connected to the objects in the exhibition", one test participant said, and another: "I have some kind of expectation at some points I'll be looking at the exhibition, but my focus was only caught on the phone". Secondly, many participants also found the experience too trivial, lacking in deeper reflection and learning. Furthermore, several participants felt unsure about the "tone" of the app: ‘I was confused and I didn't know if it was meant to be serious or not.’ Some of those who tried to use the app to express earnest, serious messages were worried that this might make them look silly: ‘my slogan was “stand for new humanity”, and “I fight against mass-manipulation”, and those kinds of things. Sort of for me, like I look silly, [but] it was serious for me’. One participant suggested that the app should make it clear to users that joking is acceptable, by including an explicit instruction in the beginning to ‘remember to have fun.’

\section{Monuments for a Departed Future}

Where Twitto was designed to be a humorous play on the narrative of the museum and closely connected to its objects, Monuments for a Departed Future (or Monuments for short) was designed as a playfully poetic and intimate experience, which focused on objects that were not, and could not be, on display in the museum. The selected objects were the ‘Spomeniks’: socialist monuments placed all over the former Yugoslavia which are not represented in the permanent exhibition. Just as with Tito himself, the monuments are focal points for ideological battles and offer rich possibilities for contrasting interpretations. This, in combination with their distinct aesthetic style, made them ideal as a critical point of entry to the history of Yugoslavia.

Monuments was designed by one of the participating researchers during a period from March to June 2017. The design built on Witcomb’s \cite{witcomb_understanding_2013} ideas on how to present visitors with “poetic provocations” in order to encourage critical engagement with “difficult” historical topics. The design goal was not only to provide an alternative historical narrative to the one provided by the museum; it was also to disrupt the ways in which visitors related to the museum and its existing content. Therefore, during the design process, it became important to develop ways in which the design could foster both a playful and an introspective mindset in its users. The aim was to trigger visitors’ imagination, build attentiveness, evoke emotions, as well as facilitate reflection.

In the final version of the game, Artcode markers graphically similar to the real monuments were used in order to give them a physical anchor point in the museum. These were placed inside the existing exhibition (Figure 4). The game provided clues on how to find each marker, inviting a playful activity of searching for the markers inside the museum and to let the hidden placement of the markers mirror how many of the monuments are located in remote locations, hidden from public consciousness. Each marker served as an entry point to one of the existing monuments as well as to a specific theme relating to their history (Figure 5). The app included eight such themes that were identified in collaboration with museum curators. After scanning a marker, the user was presented with an image of the monument and a short text on the theme. At this point the user could choose to get more historical facts about the monument, take on a challenge or answer a question. The challenges would prompt the participant to put themselves in a specific state of mind, using their imagination and their body to interact with the museum environment (Figure 6). While some of the challenges were light-hearted and playful, others more emotionally challenging. The questions were deliberately formulated to provoke reflection on contested heritage and the cultural significance of monuments. Together with the challenges, they also served to connect the experience at the museum with the user’s personal life outside of it. After submitting an answer to a question, users could view answers from other participants.

\begin{figure}[h]
  \centering
  \includegraphics[width=0.6\linewidth]{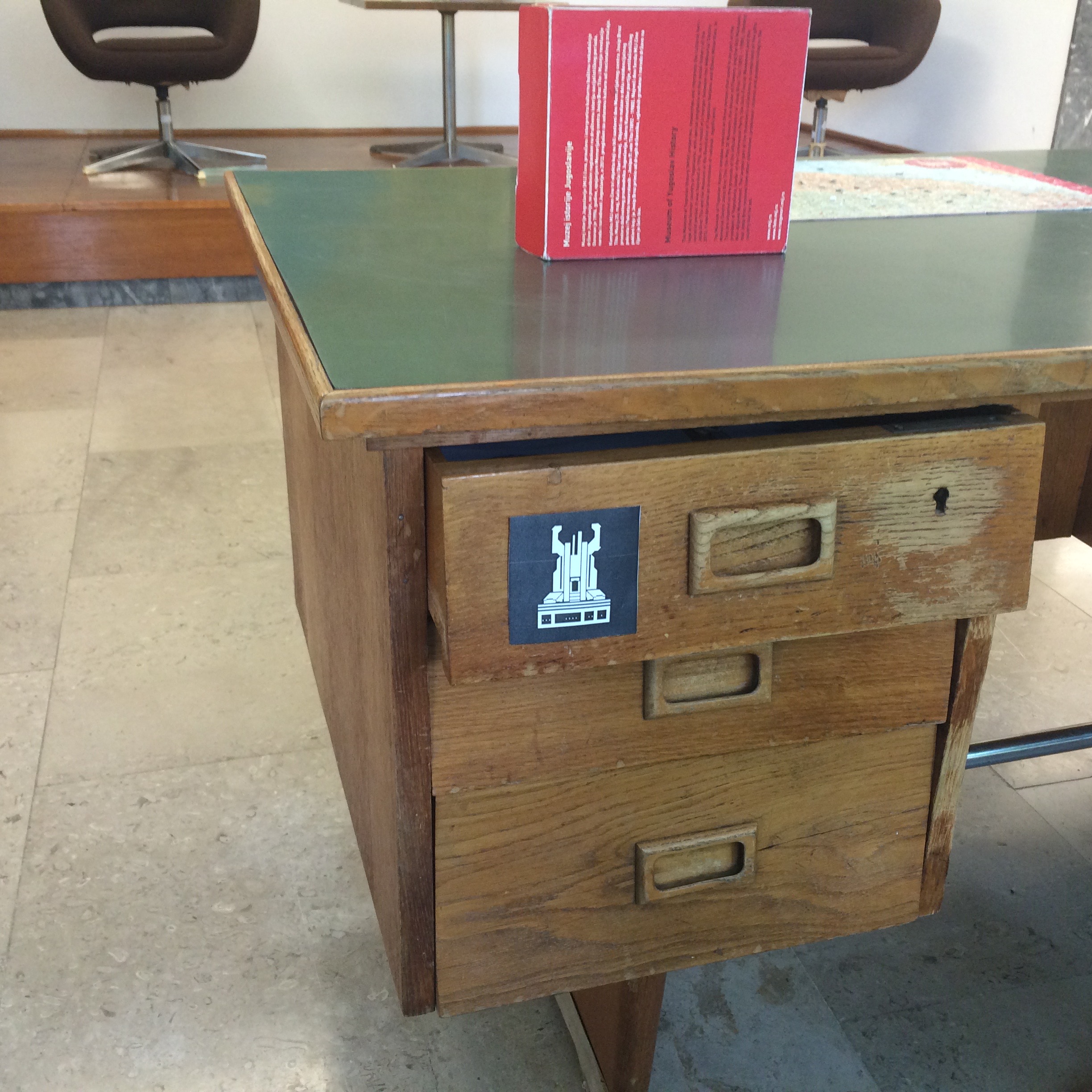}
  \caption{An Artcode representing a monument, placed at the back of a desk.}
  \Description{An Artcode representing a monument, placed at the back of a desk.}
\end{figure}

\begin{figure}[h]
  \centering
  \includegraphics[width=0.6\linewidth]{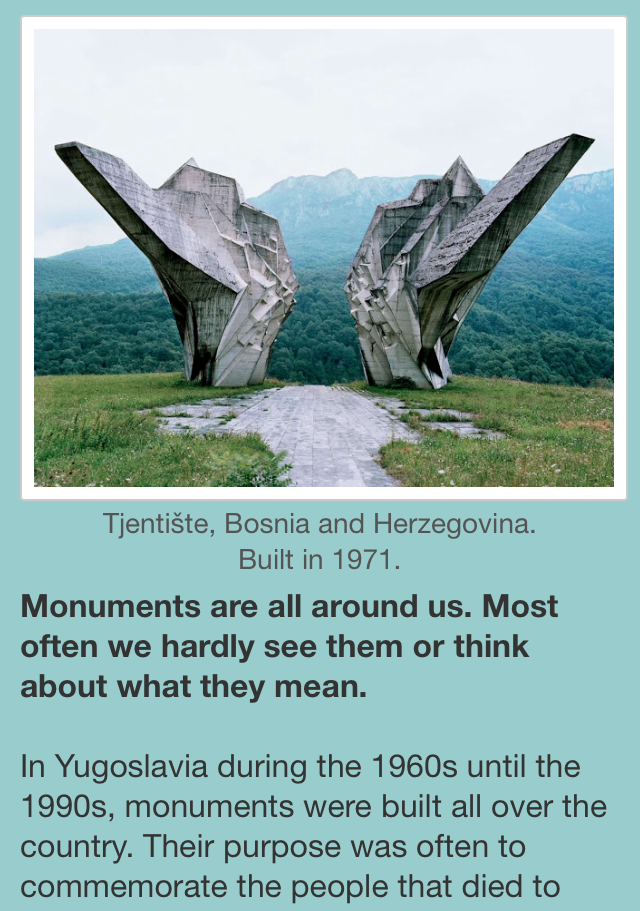}
  \caption{Screenshot from Monuments, with info snippet about one of the monuments.}
  \Description{Screenshot from Monuments, with info snippet about one of the monuments.}
\end{figure}

\begin{figure}[h]
  \centering
  \includegraphics[width=0.6\linewidth]{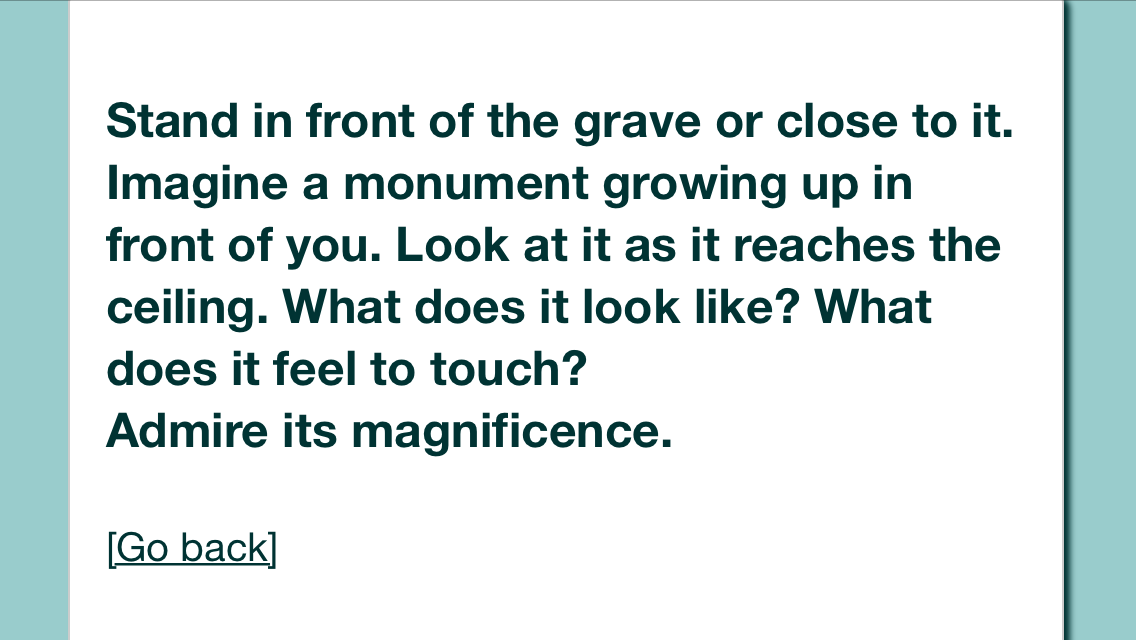}
  \caption{A poetic provocation from the Monuments app.}
  \Description{Screenshot from Monuments.}
\end{figure}

\begin{figure}[h]
  \centering
  \includegraphics[width=0.3\linewidth]{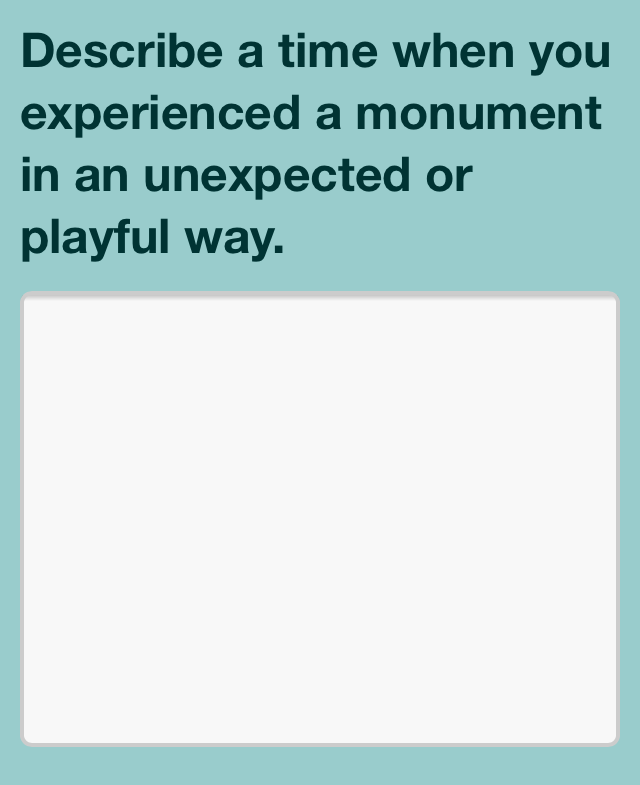}
   \includegraphics[width=0.3\linewidth]{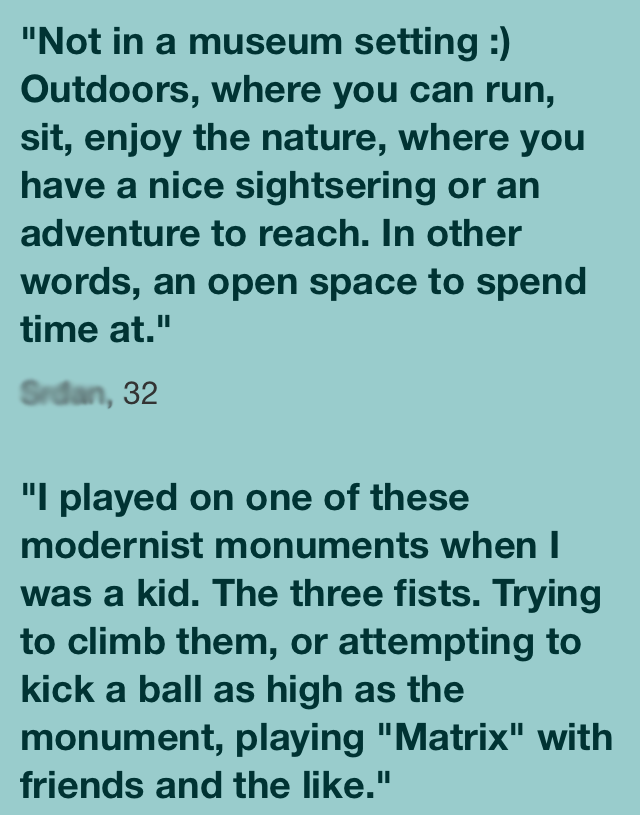}
  \caption{Question and answers from the Monuments app.}
  \Description{Screenshot from Monuments.}
\end{figure} 

\subsection{Evaluation}
Monuments for a Departed Future was implemented and trialled at the museum in June 2017; first by four art history students from the University of Belgrade (two females and two males in their early 20s) and a week later by an expert panel of nine participants (four female, five male) from the museum and the research project. The collected data from the trials is based on observations, individual interviews and a group discussion with the expert panel. 

The overall reception of Monuments was largely positive, being described as ‘spiritual’, ‘powerful’, and having a ‘thoughtful resonance’. Differently from Twitto, in Monuments players tended to take a more serious stance and be more emotionally engaged with the content. One of the participants described it as a ‘historical/emotional roller-coaster’, because of how it connected historical topics with personal life outside the museum. Another participant described it as: ‘waking you up in a way’. However, having a playful approach to sometimes very intimate and personal topics was off-putting to one participants: "One of the things that threw me off was the challenge about personal conflict and forgiveness, since I felt it was making light of that topic and I didn't like it." On the other hand, the same challenge was perceived as appropriate by another participant: ‘especially this part with the forgiveness. For me it was really deep and maybe most powerful’.

The co-created content - the answers that the participants submitted to the questions in the app - also contrasts with that of Twitto, as the answers tended mostly to reveal earnest reflections on the app's theme. For instance, responding to the question “What monuments are missing in the world today?” we find answers such as:
\begin{itemize}
\item “Monuments to love and responsibility to other people.”
\item “We need more dancing, music fountains.”
\item “Ones that represent joy and achievement and not only commemorate misery and disaster.”
\end{itemize}
However, also in Monuments some of the answers were more humorous or ambiguous, such as the following answers to the question “If these monuments were antennas to outer space, what message would you send to anyone listening?”:
\begin{itemize}
\item “Don’t come here. Humans are fucking headcases. Put a quarantine around the planet saying: ‘danger! Leave well alone!'”
\item “Who ordered the veal cutlet?”
\end{itemize}
Test players of Monuments also commented that the experience tended to direct their attention away from the physical artefacts on display, leading them rather to focus on their mobile devices. However, this reaction was much less pronounced and widespread than in the case of Twitto. 

\section{Discussion: Challenges for critical play in the museum}

The two experiences presented in this article complement each other in at least two ways, relating to the concepts of critical play and hybrid museum experiences: First, they offered quite different invitations to play – one being quite irreverent and satirical, the other being more intimate and poetic. Second, the designs related to the physical collections in different manners: One used physical artefacts on display in the museum as starting points for playful challenges, while the other focused on artefacts that were not present in the museum, attempting to give them a virtual/hybrid presence.

As noted above, the invitation to play in the Museum of Yugoslavia was challenging for both visitors as well as the museum curators. Even though both games aimed to tread a careful balance between the playful, the critical, and the historical narratives, several tensions came to the surface as they were trialled at the museum. Here we discuss issues emerging out of a) using mobile technology to foster play in the museum (as opposed to stationary interactive installations), b) the different approaches taken by the two games in terms of connecting with the exhibited artefacts and fostering critical awareness.

\subsection{Challenging norms of behaviour}
First of all, inviting visitors to play a pervasive game in the museum may easily challenge some of the visitor's expectations about norms for moving and behaving in a museum. While many museum professionals are eager to see museum spaces as open and available to a wide range of activities - including games and play - visitors come to museums with a set of norms and expectations, and may not always feel certain about which kinds of playful behaviour is acceptable. This is echoed in the study referred to earlier by Wakkary and Hatala \cite{wakkary_situated_2007}, in which they found that even though playfulness was identified positively in all aspects of their interface, the overall satisfaction was split between those participants who enjoyed playing and those who did not.  

During the play tests of Monuments and Twitto, participants could be seen looking at their phones, posing for playful selfies, searching for monument codes or sitting down in deep thought. In post-experience interviews from both games it became clear that participants were unsure whether this kind of behaviour was acceptable in the museum. Some instructions were seen as more challenging than others, as in this example from Monuments:

\begin{quote}
One of the challenges required you to close your eyes. The sitting down is all right, and it does engage you to actually go through space and do something you would not actually do. (...) But the eye-closing one… I did it but not really sincerely. (...) Just the thought of it, looking weird to onlookers was off-putting.
\end{quote}

A practical challenge also emerged when it came to fitting the hybrid experience within the time-frame of a museum visit. Some of the participants saw Monuments as a deep, contemplative experience that required a slower tempo than the one in which they would usually go through a museum. As such, they would have liked to have more time to explore the content more in depth: "I was trying to achieve it as fast as I can (...) but I would actually like to add more space to sit around just diving into what am I reading. Maybe reflecting a bit more". When it came to Twitto, several participants pointed out that the app experience left the visitor with little time to take in the exhibition. Some expressed this as feeling a bit guilty for missing out on the museum:

\begin{quote}
I felt the pressure from the game to go faster to scan the other stuff and see what's going on and I didn't take the time to look at all the stuff that I would otherwise (...) this way it was just okay, let's just scan the next thing and finish it.
\end{quote}

 Both the guilt and feelings of awkwardness expressed by the participants point to how the introduction of play in a museum,particularly when the area for play is not spatially contained or temporally limited, may serve as a disruption to the very idea of what a museum visit is. Within museology museums are sometimes referred to as ritual environments \cite{bell_making_2002, duncan_art_1995}. According to Carol Duncan, museums guide and give cues to visitors on how to perform and how to respond to the exhibits \cite{duncan_art_1995}. Even though people continually “misread”, scramble or resist cues on how to behave, most of us tend to act in a similar manner. This is confirmed by ethnographic observations \cite{veron_ethnographie_1989} as well as in quantitative studies \cite{zancanaro_analyzing_2007}. One of the characteristics of critical play, as described earlier in the article, is the transgression of norms. This is part of the transformative potential of critical play, however it is also what makes it particularly challenging to participate in for certain individuals in certain contexts. 
 
 In order to increase the number of visitors who will feel safe enough to play, social contracts need to be properly established; both between players and other visitors, as well as between players and the museum. This is always challenging when it comes to hybrid experiences, such as pervasive games, where the digital content acts as a hidden layer only accessible to those engaged in it. However, it shows how important it is both to communicate to visitors beforehand what they can expect from the experience, as well as to provide players with the possibility to opt-out at any moment. 
 
\subsection{Challenging the role of the artefact}
As described earlier, one of the central challenges in the design process of Twitto was establishing an adequate connection between the physical artefacts and the digital experience. Both museum representatives and visitors expressed a desire to avoid letting the digital device take too much of the user’s attention and divert their attention away from the physical exhibition, reflecting the common concern about the "heads-down phenomenon" described above (see section 2.2).

The design team explored multiple ways to connect the physical exhibition to digital experiences technically, narratively and through the design of visitor challenges. In Twitto, each challenge in the game took a specific object in the exhibition as its starting point, and connected the object and its significance explicitly to the topic of the challenge. For instance, scanning the Artcode next to a historic poster (see Figure \ref{wanted}) would take the player to a short series of screens explaining the significance of that particular poster, before presenting the player with the challenge to create their own propaganda poster. Still, the disconnect between physical artefacts and digital experience became a main point of criticism. It was voiced both by those participants who stated they liked the experience overall as well as by those who didn’t. The opinion was also shared among those who worked in museums and those who did not. A possible explanation may be that the great emphasis on playful co-creation quickly led the participants' attention away from the artefacts, instead getting caught up in the challenge of coming up with playful responses. Once the participant had scanned the Artcode, there was nothing in the app directing their attention back towards the physical artefact.

An alternative approach towards reorienting visitor focus towards artefacts is to design visitor tasks that emphasize the artefacts and the museum collection through the way the visitor is asked to engage with them, both narratively and through visual and bodily orientation. One of the challenges in Twitto used this approach to connect the digital with the objects on display. This challenge was associated with the part of the exhibition that dealt with Tito’s role as a resistance leader during World War II. On display is the 'Wanted' poster that can be seen in Figure \ref{wanted}, which was put up by the German forces in 1943, offering a large sum of money as a reward to anyone who gave them information leading to the capture of Tito. An Artcode placed next to this poster triggered a chapter briefly presenting Tito’s role as a resistance leader, and then tasked players with creating `Wanted' posters for themselves. They were asked to look around the physical exhibition and take a picture of a valuable object they would offer as a reward for their capture. Since the majority of the artefacts on display in this part of the museum are objects given to Tito as diplomatic gifts, the challenge was well aligned with the theme of the exhibition and there was a rich selection of items for participants to choose from. Several trial participants mentioned this challenge in particular as triggering creative ideas and experiences, in searching the exhibition for precious objects they could offer as a reward for their own capture.

In Monuments, the graphically significant Artcodes were placed on or near objects in the museum, and sometimes the playful challenges related specifically to nearby artefacts, such as Tito's grave (see Figure 6). However, for the most part the game did not address displayed artefacts, nor did it present information related to the objects in the vicinity, dealing instead with the (distant) ‘Spomeniks’. Hence, Monuments was deliberately designed to introduce a certain disconnect between physical artefacts and the digital experience in order to give presence to objects that were not in the museum. This created an experience that was, to a large degree, detached from the physical reality of the museum, turning it instead into as a stage for an ambiguous play with time and space. However, perhaps surprisingly, this did not cause the same negative reaction from test players regarding the connection to artefacts. Rather, most of the test participants were intrigued:

\begin{quote}
it's taking you somewhere else. It is taking you to the past. It is taking you to the locations where these monuments are, which are all outdoors and then you are indoors. So, you try to imagine it a bit. It's a play of spaces, which we are surrounded with.
\end{quote}

Another test player, who was already quite familiar with the museum, stated that the app made them see the museum with new eyes:

\begin{quote}
It's like I am here for the first time. It is completely changing your perspective. You really feel like you never have been here, and actually the things you're seeing now you see with completely other eyes.
\end{quote}

The difference between Twitto and Monuments regarding the participants' comments about the connection to artefacts is perhaps puzzling, given that Twitto related more closely to the physical exhibits than Monuments did. It is possible that the difference in tempo between the two games has contributed: Perhaps the time pressure described by Twitto test players led them to ignore the artefacts, instead only focusing on finding the scannable markers to advance in the game. However it seems more likely that difference was caused by the different strategies taken by the two games: In contrast to Twitto, which tried but failed to connect fully with the artefacts, Monuments may have seemed more acceptable because it was not perceived as dealing with the physical exhibits at all, but rather introducing and exploring a theme from outside the exhibition - arguably adding a relevant perspective to the exhibition, rather than detracting from it.

\subsection{Challenging the curatorial authority}
An important concern in any design process is the involvement of stakeholders. As described above, in the case of Twitto museum representatives were invited into the design process at an early stage and participated in multiple workshops and tests throughout the process. Even so, later meetings and discussions with museum representatives have made it clear that they did not feel sufficiently involved in the design. On the one hand, this is a problem because they are uncomfortable with the way in which the game presents the museum’s collections – in particular, the museum would like to reduce the strong association between Tito and the museum, instead focusing on the importance of Yugoslav history more broadly. As such, the Twitto concept with its strong focus on Tito as a historic (and satirical) figure runs contrary to the museum’s strategy. On the other hand, the nature of the game may also run counter to the professional identity of the museum professionals, as the game offers fairly little space for factual information about the collections, instead emphasizing the invitation to play and create content.

Museums have engaged in activities resembling critical play in the past. However, such activities have tended to be framed as events, often as part of interactive art performances, festivals or similar (e.g. \cite{geismar_art_2015, mannheim_dialogue_1995, lehrer_curatorial_2016}). In contrast, Twitto was designed not to form part of a special event, but rather as a permanent part of the regular exhibition, filling a conceptual role similar to that of an interactive guide application. This may have affected the expectations both of the participants and the museum professionals, leading them to expect the game to speak with the museum's "voice" as a part of the museum's official communication towards visitors. 

In a later workshop taking place after the Twitto playtest, in which the design team and museum representatives were exploring paths forward, a curator explained:

\begin{quote}
we had the same impression that the way of engagement is really interesting (...) and it was the content that was bothering us, that was the thing, we felt that it would be much better if the games would be accompanied with the knowledge of the curators, and combining the two.
\end{quote}

In Monuments, by contrast, the strong thematic focus on artefacts that were not physically represented in the exhibition may have given the experience a different framing in the participants' minds, resembling less a guide and more a thematic exhibition or event, which regularly occurs in many museums (including the Museum of Yugoslavia) and does not always have a direct connection to the museum's own collection of artefacts. 

Moreover, Monuments seems to have followed a more accepted path in terms of museum pedagogy. Using Witcomb-inspired "poetic provocations" to foster critical awareness, proved, in this case, to be a more acceptable design strategy than using satire for the same purpose. This is interesting considering that research in psychology shows that using humor can serve as a coping mechanism to deal with challenging topics and traumas\cite{wilkins2009humor, overholser1992sense, martin2018psychology}.

One might consider that the playful format not only sets up a tension between the desire to facilitate engagement, on the one hand, and to educate on the other; but there is also a potential threat to the professional roles of museum curators and educators, who might see their role as diminished when traditional modes of dissemination and education are replaced with play. In this case, this threat was, perhaps not surprisingly, even amplified by the introduction of play which encouraged visitors to turn a critical eye towards the museum itself.

\section{Concluding remarks}

In this article we have explored two prototypes of hybrid museum experiences designed to facilitate critical play, identifying some challenges relating to norms for museum behaviour, the role of artefacts and the role of curators. In conclusion we outline some suggestions for future research.

First, regarding norms for museum visits, we have noted above that the fact that our designs did not set up a sufficient 'magic circle' for play, may have contributed to the awkwardness experienced by many players. This observation could lead to two opposite strategies in future work: First, one might attempt to help future players by designing games that are more clearly positioned within spatial and temporal limits, perhaps as part of special events, in which a clear license to engage in play is established. Or, second, the challenge regarding norms might be treated as a design challenge: Can we design hybrid museum experiences that introduce elements of critical play in an "ordinary" museum visit, making playfulness an integral part of the museum visit? Another prototype from the GIFT project might serve as an example of this strategy, in which players were tasked with controlling each others as avatars during a museum visit \cite{ryding_silent_2020,ryding_play_2020}. When playtesting this prototype, some players had fun with purposely creating situations where the "avatar" would have to do strange acts in front of other visitors who were not aware that they were playing a game - thus playing with transgressing the norms of visitor behaviour.

Furthermore, it is worth noting that the ongoing COVID-19 pandemic has already lead to some changes in the ways visitors are able to engage with technology in museums. In some cases museums have had to reconsider what types of installations may be used, due to new rules for hygiene including frequent disinfection etc. One of the participating museums in the GIFT project report that visitors appear to be more willing to use interactive guides on their own smartphone devices than before the pandemic started - perhaps due to hygiene concerns with borrowing equipment from the museum. It is possible that such changes may shift visitors' habits and expectations also in the longer term, perhaps increasing the interest in smartphone-based games and playful experiences.

Similarly, one can envisage two alternative strategies for future work in addressing the challenge with connecting hybrid experiences to physical artefacts: On the one hand, future research might explore alternative approaches that allow digital interactions to serve as an augmentation but not a distraction from the physical artefacts. One promising avenue for such research are designs based on image recognition, as employed in the popular museum app Smartify \cite{noauthor_smartify_2019}, which can recognize artworks that the user point their camera at, and returns information about that artwork. By focusing the user's interaction on the artwork itself, rather than a scannable marker, this technique may help secure that the user's attention stays with the artefact. A central challenge for such work is designing the information delivery and interactions that happen after the artwork has been scanned. One approach to this challenge is explored in \cite{lovlie_designing_nodate}.

On the other hand, future work might choose to challenge the notion that museum artefacts should always be the centre of attention for museum visitors, reflecting the museological debate about whether museums are primarily about objects or ideas. The design of \textit{Monuments for a Departed Future} points out one possible direction for such designs. However, designers and researchers going in this direction should be prepared for critical reactions from museum curators.

Challenging the role of curatorial authority may in some regard be unavoidable, as creating hybrid experiences requires that the museum involves outside experts such as designers and developers. This entails that the museum professionals hand over some of the power (and responsibility) of shaping the museum experience. However, this does not necessarily mean that they are willing to hand over also the power (or responsibility) for the content and curatorial decisions involved in the experience - even if this runs counter to the ideals of new museology and critical museology, which often emphasise relinquishing control over curation and favoring more participatory approaches. Given the playful implications of hybrid experiences, the borderline between design and curation may be challenging to navigate. This means that designers of hybrid museum experiences which foster play - especially critical play - may need to pay much attention to involving curators in the process, and negotiating responsibilities regarding content and presentation. Similarly, museum professionals interested in facilitating critical play through hybrid experiences should be prepared to encounter multiple challenges, and perhaps may need to let go of more of their curatorial control than they might expect.

\section{Acknowledgments}
This project has received funding from the European Union’s Horizon 2020 research and innovation programme under grant agreement No 727040 (the GIFT project).

\bibliographystyle{ACM-Reference-Format}
\bibliography{references}

\end{document}